\documentclass[
    aip,
    jcp,
    amsmath,
    amssymb,
    preprint
]{revtex4-2}

\usepackage{graphicx}   
\usepackage{dcolumn}    
\usepackage{bm}         
\usepackage{amsfonts}
\usepackage{amsthm}
\usepackage{xcolor}
\usepackage[percent]{overpic}
\usepackage{comment}

\begin{document}

\title{Efficient Searches for Low-Energy Structures in Clusters with
Thousands of Particles: Application to the Thomson Problem}

\author{Paolo Amore}
\email{paolo@ucol.mx}
\affiliation{Facultad de Ciencias, CUICBAS, Universidad de Colima,
Bernal D\'iaz del Castillo 340, Colima, Colima, Mexico}

\author{David J.~Wales}
\email{dw34@cam.ac.uk}
\affiliation{Yusuf Hamied Department of Chemistry, University of Cambridge,
Lensfield Road, Cambridge CB2 1EW, UK}


\begin{abstract}
We propose an algorithm for identifying low-energy minima in systems
containing thousands of particles, at moderate computational cost. We focus
on spherical crystals, in which particles form ordered structures on the
surface of a sphere. Our method uses symmetric seed configurations, or known
structures, to initialize searches across contiguous ranges of system sizes.
Newly identified minima are then used to guide searches at neighbouring sizes,
allowing information to propagate efficiently through configuration space.
We have tested this approach on the Thomson problem, one of the few
interacting-particle systems for which global optimization has been considered
for systems of this size. We substantially improve on many previous solutions,
particularly at larger system sizes, where systematic global optimization is
computationally expensive. Since favourable packings for the Thomson problem
are reflected in systems spanning atomistic to mesoscopic length scales, the
newly characterized defect patterns may inform structure prediction for a
broad range of problems with spherical topology.
\end{abstract}


\maketitle

\section{Introduction}
\label{intro}

J.~J.~Thomson first proposed the question of how a set of like charges arrange on a sphere in the context of atomic structure \cite{Thomson04}.
Locating the lowest energy state for $N$ identical charges confined to a spherical surface is now known as the Thomson Problem.
The solutions do not provide insight into atomic structure, but have found widespread applications for systems with spherical topology,
which cover a remarkable range of length scales. 
Examples include virus capsids \cite{CasparK62,MarzecD93}, colloids \cite{disc03,LipowskyBMNB05,EinertLSBB05,ChoYLKMPY05},
fullerenes \cite{krotohocs85,fullerene00}, lipid rafts \cite{SimonsV04}, and superconducting films \cite{superconductor,vortice}.
The `Distribution of Points on the 2D-Sphere' also features at number seven in Smale's list of eighteen unsolved problems 
in mathematics for the 21st century \cite{Smale1998}.
Analytical solutions for the global minimum are only available in the small size regime \cite{Schwartz2013, Schwartz2020}.
However, the long range of the Coulomb potential means that there are fewer local minima than for atomic clusters or molecules with
a comparable number of particles \cite{doyewb95,doyew96d,doyew97a},
which makes numerical investigations more tractable.
Nevertheless, the number of local minima for $N$ like charges is expected to grow exponentially with size \cite{stillingerw84,ErberH95,WalesD03},
so that larger systems rapidly become challenging for any global optimisation procedure
\cite{siman86,ErberH91,Bowick,%
constrained,ErberH95,altschuler97,ErberH97,genetic,altschuler05,AltschulerP06,WalesU06,WalesMA09,Amore25,Amore26}.

In view of the insight obtained from the Thomson Problem in such a wide range of applications, investigating
low energy solutions is an interesting field. 
Euler's theorem for the spherical packing requires a topological
or disclination charge of 12, where $Q=6-C$ is the topological charge for a particle with $C$ neighbours.
The evolution of different packing defects with $N$ to satisfy the Euler condition is insightful,
since we expect mechanical and transport properties to depend on how defects respond to external perturbations.
For small systems the lowest minima often exhibit twelve particles with five neighbours, corresponding to disclinations \cite{livshits99,BowickCNT02}.
However, dislocations with adjacent five-coordinate
and seven-coordinate particles emerge for larger sizes \cite{Nelson02,disloc99,fullerene00},
and features that resemble grain boundaries appear as $N$ increases further \cite{BowickCNT02,disc03,BowickCNT06}.
Predicting the most favourable defect motifs and how they can interconvert may provide insight into observable properties;
for example, dislocations feature in studies of mechanical response in carbon nanotubes \cite{DumitricaHY06},
expansion of bacteria \cite{HarrisS70}, and cell division \cite{Harris75}.
The organisation of the corresponding energy landscape encodes information about rearrangement pathways and emergent
thermodynamic and kinetic properties.
Small world characteristics have been identified for the Thomson problem, with local minima separated from the global 
minimum by relatively short pathways \cite{Mehta2016}.
This organisation is often associated with efficient self-organisation in terms of relaxation to the global minimum
\cite{LeopoldMO92,BryngelsonW87,onuchicwls95,WalesMW98}.

The problem itself is easily stated, namely to find the lowest energy for the potential energy of $N$ unit charges 
on a sphere:
\begin{equation}
\label{eq:Thomson}
V=\sum_{i<j}^N \frac{1}{|{\bf r}_i-{\bf r}_j|^s},
\end{equation}
with $|{\bf r}_\alpha|=1$ for all $\alpha$ and {\color{black} $s=1$ for a Coulomb potential}.
In the present contribution we report minima for the Thomson problem,  which improve upon previous 
solutions, using a combination of step-taking and minimisation approaches that together can be considered
a form of basin-hopping global optimisation \cite{lis87,walesd97a,waless99}.
In particular, we employ moves between adjacent system sizes, which were exploited in early basin-hopping
applications to atomic clusters \cite{walesd97a} and recent applications to the Thomson Problem \cite{Amore25,Amore26},
along with symmetry considerations.
Steps between different cluster sizes and symmetry are often exploited in global optimisation; 
here we have developed a systematic approach that exploits various features of the Thomson problem
to provide a computationally inexpensive way to reach low-lying structures quickly.
Configurations with a greater symmetry content are expected to have particularly high or low energy \cite{Wales03},
meaning that low-lying solutions are more likely to have configurations with higher or partially broken point group symmetry.
If approximate symmetry is manifested in terms of near degeneracies, then correlation is expected within the contributions that
sum to give the total energy.
This observation suggests that structures with a higher symmetry content would follow a distribution of energies 
that exhibit higher variance, containing members with particularly high and particularly low energy \cite{OakleyJW13}.
Analogous connections have been identified for symmetry and designability of heteropolymers,\cite{lihtw96,wolynes96,LindgardB96}
and for mutational effects\cite{NelsonTO97,NelsonO97}.

The basin-hopping approach simply involves steps of some kind between local minima, which remove downhill barriers from the
landscape, and an accept/reject criterion based on properties of these stationary points. 
Numerous variants have been used,\cite{LearyD99,Goedecker04,SchoenbornSE2009Tpom}
including alternative accept/reject statistics,\cite{ShangW14}
and memory of previous visits.\cite{Wales10c}
Generalised basin-hopping schemes were originally developed for nanoalloys,\cite{SchebarchovW13,SchebarchovW14}
with mutiminima defined in different metric spaces,
and have subsequently been exploited in quantum computing to optimise wavefunctions based on 
discrete operator ordering and continuous amplitudes,\cite{BurtonMTW23}
and for mutational optimisation of biomolecular sequences.\cite{RoederW18b}
The objective function is usually a potential energy, but local measures of free energy,\cite{SutherlandCashWC15,MantellPW16,SutherlandCashMW17}
and grand or semigrand potentials,\cite{CalvoSW16} have also been used.
Our current scheme exploits propagation of favourable solutions between sizes, sweeping successively up and down through $N$.
In this sense there is some resemblance to grand canonical basin-hopping, but we do not employ a chemical potential 
in the present work.
Rather than using systematic random coordinate perturbations to propose moves, we reduce the computational
expense with targeted steps.
Since a variety of alternative strategies are employed, we therefore refer to this approach as
Oscillating Size  Hybrid Basin-Hopping.
Aside from interesting physical applications, studying the low-energy configurations of a large number of interacting particles on a sphere is 
also important in investigating the asymptotic behaviour of the energy at large $N$~\cite{Saff94, Saff95, Saff97, Saff19}.

\section{Oscillating Size  Hybrid Basin-Hopping}
\label{sec:algo} 

Our aim was to develop a computationally inexpensive algorithm specifically designed to identify low--energy minima 
and provide better solutions to
the Thomson Problem defined in equation (\ref{eq:Thomson}) for larger systems, by which we mean more than 1,000 charges.
The algorithm incorporates the upgrade/downgrade
approach introduced in ref.~\cite{Amore25} for the Thomson Problem and refined in
ref.~\cite{Amore26}, which uses information from adjacent system sizes, as in 
early basin-hopping investigations for atomic clusters \cite{walesd97a}.

The search is based on the following steps:
Let $N^\star$ be the number of charges of the target configuration
and consider the range $N_{\rm min} \leq N \leq N_{\rm max}$. We assume that a
low energy solution or guess is available at some $N_0$ inside this region. This guess can be achieved
by a preliminary optimization, or if results from the
literature are available. Another option
is to look for local minima under symmetry constraints
(typically point groups $T$ or $I$): given the reduced number of degrees of freedom, higher symmetry solutions can
be obtained numerically relatively quickly.

Clearly, we have three possibilities: $N_0 < N^\star$, $N_0 > N^\star$ and $N^\star=N_0$.

For $N_0 < N^\star$, the parent configuration has fewer points than the target configuration, so we need to add 
points to reach the final configuration. This addition is done in steps. 

First we consider the Voronoi diagram of the parent configuration: under the
assumption that only vertices with three edges are present, there are $2N_0-4$
Voronoi vertices. To generate a daughter configuration with $N_0+1$ points, we
consider the $2N_0-4$ 
ansatzes obtained by adding a single charge to the parent configuration corresponding to one of the Voronoi vertices and
for each of them we calculate the individual potential energy of the added
charge. These steps are computationally light, since they do not involve
minimization, but the number of ansatzes has to be reduced: in general we want
to select $p \ll 2N_0-4$ candidates.
The selection of the suitable ansatzes can be done stochastically, but we have found out that a greedy approach, consisting of
accepting the $p$ configurations with lowest individual energy is quite efficient. 

After selecting the $p$ best ansatzes, each one is minimized, landing in local
minima with $N_0+1$ particles; if a configuration with $N_0+1$ particles was
already available it is compared with the new configurations, and the overall
best configuration is kept. The process can now be iterated several times,
until reaching $N_{\rm max} \geq N$. 

Whenever better local minima are found, their structures
`propagate' to neighbouring sizes. 
We note that the modifications to 
parent configurations are quite localised, and the cost of subsequent minimization is smaller
than for an arbitrary perturbation.
Previous basin-hopping studies employed coordinate perturbations of all the particles \cite{WalesU06,WalesMA09},
which corresponds to unbiased searches where the minimisation steps are more expensive.

For $N_0 > N^\star$, the approach is similar: in this case, however,
the parent configuration has more points than the target configuration and we
need to eliminate one. This reduction is done by first considering the $N_0$ ansatzes
obtained by eliminating each of the original points from the parent
configuration and by calculating for all them 
the individual energy for each of removed charges, without relaxation.

Once again one can adopt a  greedy approach, accepting the $p$ configurations
corresponding to removing the charges with the largest individual contribution to the energy. Once the
ansatzes are determined they are minimised and the process continues
until reaching $N_{\rm min} \leq N$;

Upon reaching the values $N_{\rm min}$ or $N_{\rm max}$, the search
can bounce and reverse direction; a proper stopping criterion must also be
enforced (typically no improvements on the whole region);

In the special case $N_0 = N^\star$ one can consider both particle addition and elimination,
as described above, reversing course when reaching the
limits of the interval;

It is worth noting some technical aspects that are relevant in the
execution of the above framework: 
while there is no rigorous criterion to choose the starting $N=N_0$ and the
region $N_{\rm min} \leq  N \leq N_{\rm max}$,  an initial configuration with
reasonably low energy close to the target configuration is a natural
candidate. However, we have also discovered cases of parent configurations
with higher energy and further away from the target, which have evolved to
optimal solutions.  
For example, in the case of $N=4352$ we have used a parent
configuration at $N_0=4320$, with rather high energy, and found that it evolved
to a new lowest energy solution for this size. 

A pragmatic approach is to first explore closer parent
configurations and gradually enlarge the search to more distant configurations, if
the improvements are not satisfactory. Of course, choosing parent
configurations that are not very close (in $N$) comes with a larger
computational cost, but the effort may be rewarded
with lower energy solutions that propagate over a wide range of $N$. 

When choosing a parent configuration it may also be helpful to
look at its Voronoi diagram, and 
check the topological defect structure. Because the algorithm operates
with a sequence of localised perturbations, while moving to different sizes,
the daughter configurations may retain some of the features of the
parent configuration. 

In some cases the lowest energy solutions have $I$
or $T$ point group symmetry (see for example the Cambridge Landscape Database or ref.~\cite{WalesMA09}),
and therefore the application of the algorithm with these parent configurations
naturally produces configurations with slightly broken symmetry.
The extent to
which the symmetry is broken will both depend on both the distance from the parent
symmetrical configuration and on the number of bounces performed by the
algorithm. This observation confirms and extends the results of Roshal et
al.~\cite{Roshal15}, which improved previous records by looking for
configurations with slightly broken icosahedral symmetry (we will further
comment on this point in the section of numerical results).

Aside from these considerations, we also wish to point out that the approach that we have adopted 
is intentionally minimal,  keeping the algorithm as simple as possible. The main justification for
this choice is the success in improving previous records, particularly for configurations with the largest number of charges.

For completeness, we mention some of the features that we have
not yet fully explored, which may be worth  considering in future work: 

{\it Allowing larger steps in the number of particles.}
At present the algorithm performs unit steps in system size, generating daughter
configurations with either a point less or a point more than the parent.
However, this
procedure can obviously be extended to perform larger steps, such as from $N$ to
$N+2$.  To implement larger steps we could first calculate the
individual energies of each of the Voronoi vertices, ranking them in energy and
generating $p$ intermediate configurations corresponding to the $p$ lowest
energies, obtained by merging one of these vertices into the parent configuration.
Then the process can be repeated by calculating, for each intermediate configuration, the
individual energy of the Voronoi vertices, and this time picking the one with lowest energy. 
This way we are left with $p$ ansatzes that can then be
minimised. The passage from $N_0$ to $N_0-2$ can be done in a similar way.  
Choosing the lowest energy is analogous to threshold acceptance,\cite{HoffmannFS02}
and to downhill only basin-hopping, which has located new global minima successfully
via a multistart procedure.\cite{LearyD99}

{\it Allowing independent processes.} 
In the current implementation of the algorithm a fixed number of ansatzes (in
our case typically $32$) is generated  at each step and then minimized; the
lowest energy configuration is then retained, compared with previous record, and
kept only if it improves it.
Alternatively, one could use
independent walkers (one for each ansatz), spanning the same
region of $N$, but not necessarily at the same $N$. 

{\it Performing intermediate global optimization.}
In its present form, our algorithm starts from a parent configuration and tries
to find improved configurations
over a specified region of $N$. At each stage, any solution improving a
previous record is accepted and directly used as parent configuration in the next step. 
It may be worthwhile to supplement each step with a limited exploration of the
landscape in proximity of the improved configuration, using, basin-hopping\cite{lis87,walesd97a,waless99}
steps with random coordinate perturbations. 
This modification, however, would probably come with a significant computational cost.

Finally we mention an important technical detail: at any time, we are working with $2N-3$
degrees of freedom, since three angles are kept fixed (in our convention, we
put a point on the north pole and another in the xz plane). 
At each step, after producing the ansatzes and minimising them, the parent
configuration is rotated to move a different
pair of points to the reference position. Ideally, if we had worked with the
whole $2N$ degrees of freedom this setup would not make a difference, but since the three angles are
frozen, this approach can change the outcome of the minimization. 
A configuration previously visited by the algorithm may therefore
produce different final configurations upon minimization.

\section{Numerical Results}
\label{sec_nr}

As noted in the Introduction, Euler's theorem requires that the total topological charge of any arrangement of points on the sphere 
is $12$. For sufficiently small systems this condition is usually satisfied in the global minimum
for 12 disclinations.
However, as the number of charges increases, more complex defect patterns, still constrained by Euler’s theorem, become
more energetically favourable.
In particular, Ref.~\cite{GarridoDM97} investigated relatively large
configurations of points on the sphere and showed that the appearance of
dislocations can lower the energy, albeit at the cost of breaking icosahedral
symmetry. Complex arrangements of topological defects were also found in
symmetric configurations constructed using Toomre’s ring-removal method~\cite{disloc99}.
Finding the true global minimum of the total energy for larger systems becomes
increasingly challenging, because of the exponential growth with $N$
of the number of local minima.\cite{stillingerw84,ErberH95,WalesD03} 
The exponential scaling with $N$ is relatively favourable for the long-range Coulomb 
potential,\cite{doyewb95,doyew96d,doyew97a} but nevertheless we expect that the lowest minima 
located in the larger-size regime will be improved in the future.
However, the defect motifs that emerge may still provide insight into packing
problems for spherical topology, where accurate representations of more complex
interatomic potentials for systems such as biomolecules would make computational
searches problematical.

To assess the performance of the OSHBH algorithm 
we have considered configurations with $N \geq 500$ charges, using previous reports
\cite{Wales06,AltschulerP06,Roshal15,WalesMA09}
for comparison (a recent paper, ref.~\cite{Lai24}, by Lai and collaborators,
also considered global optimization for configurations with $N \leq 500$).
For each of these references we have performed a comparison of the numerical results, which are organize in 
tables \ref{tab:Wales06}, \ref{tab:Altschuler}, and \ref{tab:roshal}, containing 
results for configurations of intermediate size, $500 \leq N \leq 1000$, and $ N \leq 4352$, respectively.

In the first set of tests (first three tables) we see that the comparison with previous results is quite favourable, and, even when the
previous record is not reproduced, the energy of the configuration found is quite close (see Table \ref{tab:Wales06}).
Because the results of ref.~\cite{Wales06} were obtained using systematic global optimization (basin-hopping
with random coordinate perturbations),\cite{lis87,walesd97a,waless99} 
we conclude that the exploration with the current algorithm produces configurations of good quality with 
significantly less computational effort. These configurations could be used as starting points for full global optimization.

The comparisons with refs.~\cite{AltschulerP06} and \cite{Roshal15} are different,
because these studies did not perform a full global search.
Not surprisingly, we have been able to improve (significantly) on most of these results
(see tables \ref{tab:Altschuler} and \ref{tab:roshal}). It is likely that the
configurations that we have found could still be further improved upon by
applying global optimisation.
In fact, we are finding improved structures quite regularly as the algorithm continues to run.
We have therefore chosen to report the current best efforts as a snapshot of progress, and
we will automatically update the best solutions on new web pages in the
Cambridge Landscape Database\cite{CLD} so that they are publicly available.

\begin{table}[!t]
\centering
\begin{tabular}{|c|l|c|}
\hline
\textbf{$N$} & \textbf{Ref.~\cite{Wales06}} & \textbf{This work}  \\
\hline
522 & 129655.32240670 & 129655.32240673 \\
\hline
572 & 156036.21929100 & $156036.23856348^\star$ \\
\hline
612 & 178909.79524900 & $178909.79527782^\star$ \\
\hline
642 & 197097.19930020 & 197097.19930022 \\
\hline
672 & 216169.99399450 & $216170.04035469^\star$ \\
\hline
732 & 256972.43588720 & 256972.43588724 \\
\hline
752 & 271360.98891960 & 271360.98891959 \\
\hline
762 & 278703.06088150 & $278703.10859718^\star$ \\
\hline
792 & 301319.87339170 & 301319.87339173 \\
\hline
812 & 316890.45805410 & 316890.45805414 \\
\hline
842 & 340985.65763870 & 340985.65763870 \\
\hline
912 & 400657.61650270 & 400657.61650272 \\
\hline
932 & 418594.23395000 & 418594.23394996 \\
\hline
972 & 455651.08093510 & 455651.08093509 \\
\hline
\end{tabular}
\caption{Lowest energy configurations of the Thomson problem for selected values of $N$ reported in ref.~\cite{Wales06}. 
Values marked with a star are slightly above the results of \cite{Wales06}.}
\label{tab:Wales06}
\end{table}

\begin{table}[!t]
\centering
\begin{tabular}{|c|l|c|}
\hline
\textbf{$N$} & \textbf{Ref.~\cite{AltschulerP06}} & \textbf{This work}  \\
\hline                        
512 & 124674.44799833 & 124673.90556766 \\
\hline                        
516 & 126655.45229018 & 126654.60408762 \\
\hline                        
518 & 127651.79196193 & 127650.80759297 \\
\hline                        
532 & 134735.82804448 & 134735.07154548 \\
\hline                        
534 & 135763.57340729 & 135762.91066161 \\
\hline                        
538 & 137830.85765544 & 137830.13608932 \\
\hline                        
542 & 139913.59874297 & 139913.14545394 \\
\hline                        
548 & 143068.03051847 & 143067.00646283 \\
\hline                        
672 & 216171.22752456 & $216170.04035469^\star$ \\
\hline                        
722 & 249927.18054807 & 249926.16628719 \\
\hline                        
762 & 278704.42807713 & 278703.10859718 \\
\hline                        
777 & 289902.13212320 & 289900.80824298 \\
\hline
\end{tabular}
\caption{Lowest energy configurations of the Thomson problem for selected values of $N$
reported in ref.~\cite{AltschulerP06}. Our value for $N=672$ is marked with a
star because it is slightly above the value reported in the Cambridge Database.
}
\label{tab:Altschuler}
\end{table}

Roshal and coworkers advanced the interesting observation that
lower energy configurations could be 
obtained by slightly breaking the icosahedral symmetry, finding in this way a
series of configurations (all in the range $N < 1000$) that improve previous
records.\cite{Roshal15} In fact, nine out of the fourteen solutions described in the previous report have been
improved  using the present algorithm, while reproducing  the other five results (see table \ref{tab:roshal}).

\begin{table}[!t]
\centering
\begin{tabular}{|c|l|c|}
\hline
\textbf{$N$} & \textbf{Ref.~\cite{Roshal15}} & \textbf{This work}  \\
\hline                        
620 & 183673.09953200 & 183673.09953203 \\
\hline
660 & 208434.78118710 & 208434.61612885 \\
\hline
739 & 261964.00459477 & 261963.50207710 \\
\hline
740 & 262680.38954930 & 262680.38954931 \\
\hline
743 & 264837.70097130 & 264837.58585036 \\
\hline
745 & 266280.73119190 & 266280.25892291 \\
\hline
750 & 269904.22623690 & 269904.22596347 \\
\hline
753 & 272090.80855654 & 272090.80855654 \\
\hline
755 & 273553.41303680 & 273553.33173358 \\
\hline
759 & 276490.45478854 & 276490.08438220 \\
\hline
780 & 292166.76583720 & 292166.76583725 \\
\hline
900 & 390084.75568630 & 390084.75562010 \\
\hline
920 & 407784.83883230 & 407784.83883235 \\
\hline
960 & 444368.81934810 & 444368.43734798 \\
\hline
\end{tabular}
\caption{Configurations of the Thomson problem for selected values of $N$ reported in ref.~\cite{Roshal15}. }
\label{tab:roshal}
\end{table}

Ref.~\cite{WalesMA09}, on the other hand, contains the largest configurations for the Thomson problem obtained from
systematic global optimization, using a standard basin-hopping approach.\cite{lis87,walesd97a,waless99}
A comparison with the results of ref.~\cite{WalesMA09} is therefore 
essential to test the computationally cheaper approach we have defined here.
Table \ref{tab:WalesMA09} reports the results for the cases of \cite{WalesMA09} where we have been able to improve 
the energies, which essentially correspond to all cases $N \geq 2000$ considered in the previous report, with one
special case at $N=1382$. These configurations have been obtained mostly by
first locating a low energy symmetric configuration in a neighbourhood of the
targeted $N$ and performing repeated bounces over a region of $N$ that
includes the target. The computational time required for this process depends on several factors, but in
particular on the size of the configurations, ranging from a few hundred seconds
to complete each new $N$ in the bouncing for the smallest configurations, to a few
thousand seconds for the largest. The corresponding hardware is defined below, and the timings correspond to
calculations on a single core.

\begin{table}[!t]
\centering
\begin{tabular}{|c|l|c|}
\hline
\textbf{$N$} & \textbf{Ref.~\cite{WalesMA09}} & \textbf{This work}  \\
\hline
1382 & 926576.11756060 & 926576.10303295 \\
\hline
2000 & 1950575.63893870 & 1950575.58888456 \\
\hline
2192 & 2345722.29935010 & 2345720.993207238 \\
\hline
2500 & 3055924.21311050 & 3055923.109565069 \\
\hline
2502 & 3060843.53713850 & 3060842.24038072 \\
\hline
2832 & 3926826.08864710 & 3926824.75006776 \\
\hline
3182 & 4963365.82683580 & 4963364.318860015 \\
\hline
3552 & 6191357.25796330 & 6191355.686658503 \\
\hline
3942 & 7632895.02893720 & 7632893.156491106 \\
\hline
4352 & 9311276.27551950 & 9311272.990970789 \\
\hline
\end{tabular}
\caption{Lowest energy configurations of the Thomson problem for selected values of $N$ reported in ref.~\cite{WalesMA09}  }
\label{tab:WalesMA09}
\end{table}

\begin{table}[!t]
\centering
\begin{tabular}{|c|c|c|}
\hline
\textbf{$N$} & \textbf{basin-hopping (2023)} & \textbf{This work}  \\
\hline
4352 & 9311275.08620466 & 9311272.990970789 \\
\hline
5792 & 16529993.08369918 & 16529987.43916848 \\
\hline
\end{tabular}
\caption{Lowest energy configurations of the Thomson problem for $N = 4352$ and $N=5792$ obtained using basin-hopping 
with random coordinate perturbations for the moves in 2023 
(Wales, unpublished work).  }
\label{tab:Walesprivcomm}
\end{table}

The cases $N=4352$ and $N=5792$ had also been improved in the recent past using
standard basin-hopping; in both cases our method now delivers results that improve the
previous record by a substantial amount (see  table \ref{tab:Walesprivcomm}).
In particular, the calculation for $N=5792$ was identified using the lowest
energy $T$ symmetry configuration found at $N=5790$ and performing repeated
bounces in the region $5790 \leq N \leq 5794$. Eventually the symmetrical
parent configuration was also improved.

\begin{figure}[!t]
\begin{center}
\includegraphics[width=12cm]{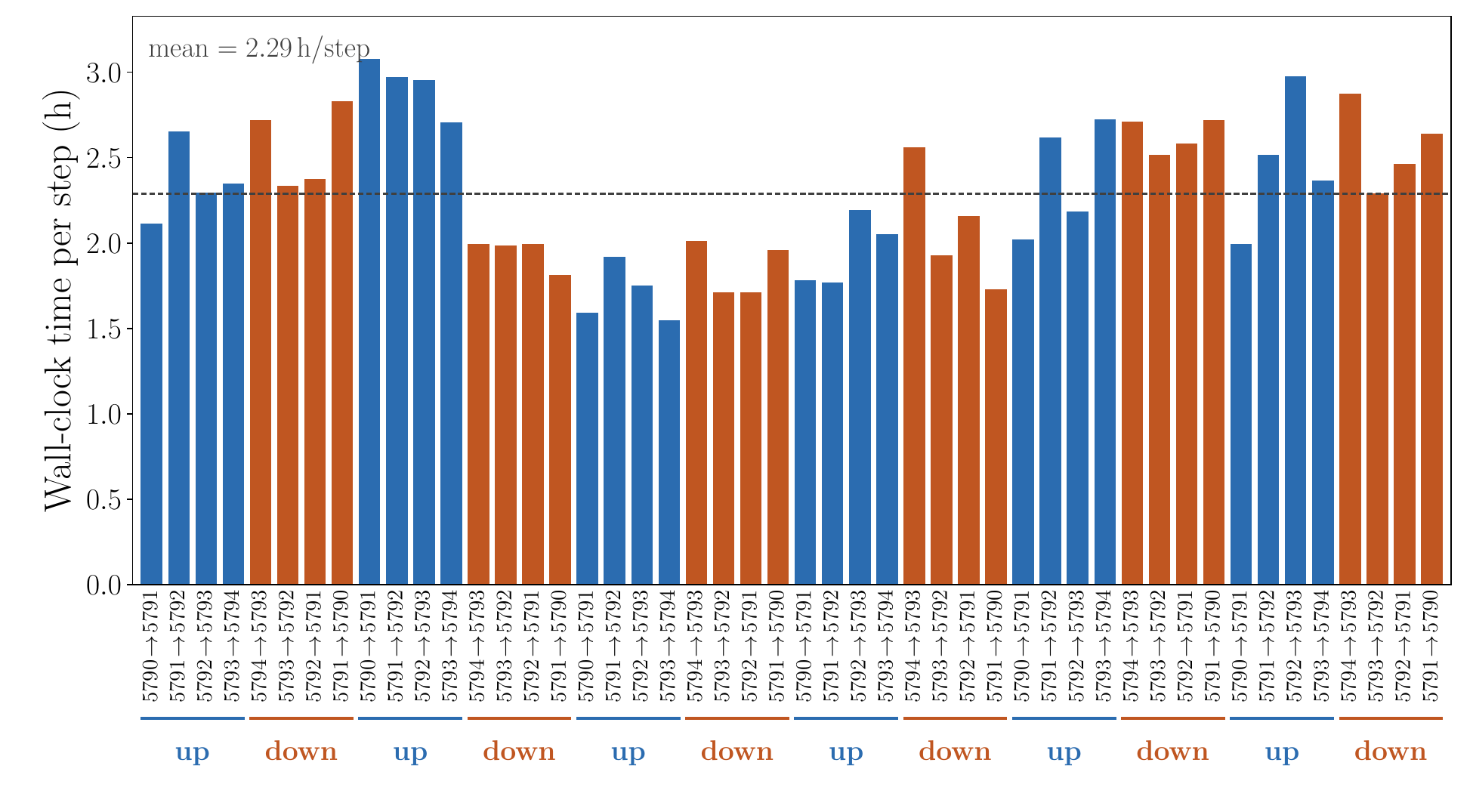}
\caption{Timings for steps involved in the bouncing calculations around $N=5792$.}
\label{Fig_timing}
\end{center}
\end{figure}

In Fig.~\ref{Fig_timing} we report the timings for the steps performed by the
algorithm, either in the upward (blue) or downward (brown) direction; the runs
were performed on a Linux (ubuntu) computer with an AMD Ryzen 9 7950X, running
$32$ processes in parallel at all times. Each step required an average of $2.29$\,hours to complete. 
No appreciable timing difference is observed between up and down passes.

\begin{figure}[!t]
\begin{center}
\includegraphics[width=12cm]{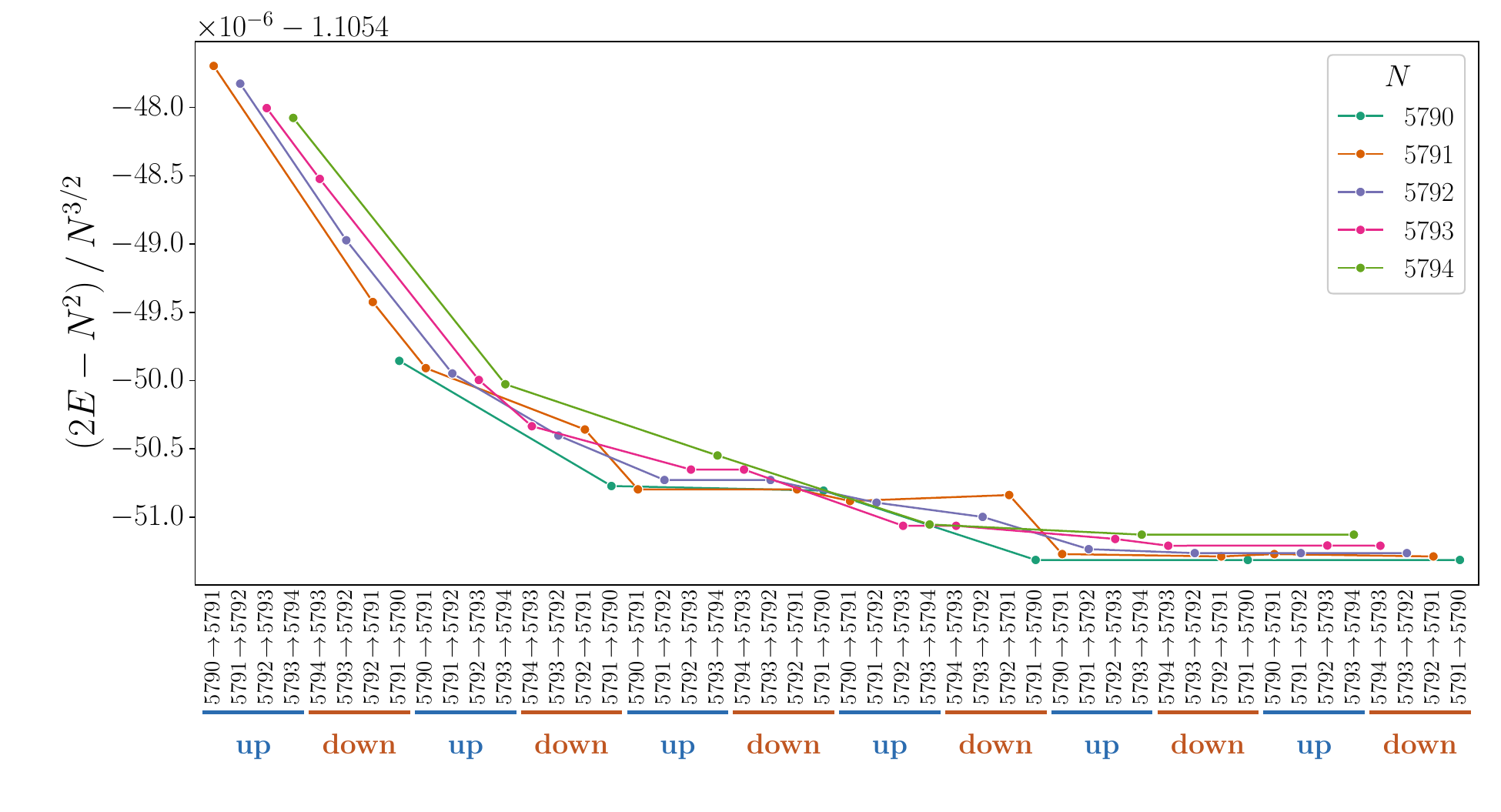}
\caption{Scaled energies $(2 E_N-N^2)/N^{3/2}$ reached in the bouncing about $N=5792$.}
\label{Fig_energies}
\end{center}
\end{figure}

In Fig.~\ref{Fig_energies}  we plot the scaled energies $(2 E_N-N^2)/N^{3/2}$
reached by the algorithm. Taking out the continuum contribution
$N^2/2$ provides a more interpretable comparison between configurations
of different $N$.
As one can see from this plot,  there is a larger drop of the energy in the
first stages of the algorithm, until reaching a plateau in the last stages (the
stopping criterion is that no improvement is seen in a complete sweep of the $N$ region).

\begin{figure}[!t]
\begin{center}
\bigskip
\includegraphics[width=3.95cm]{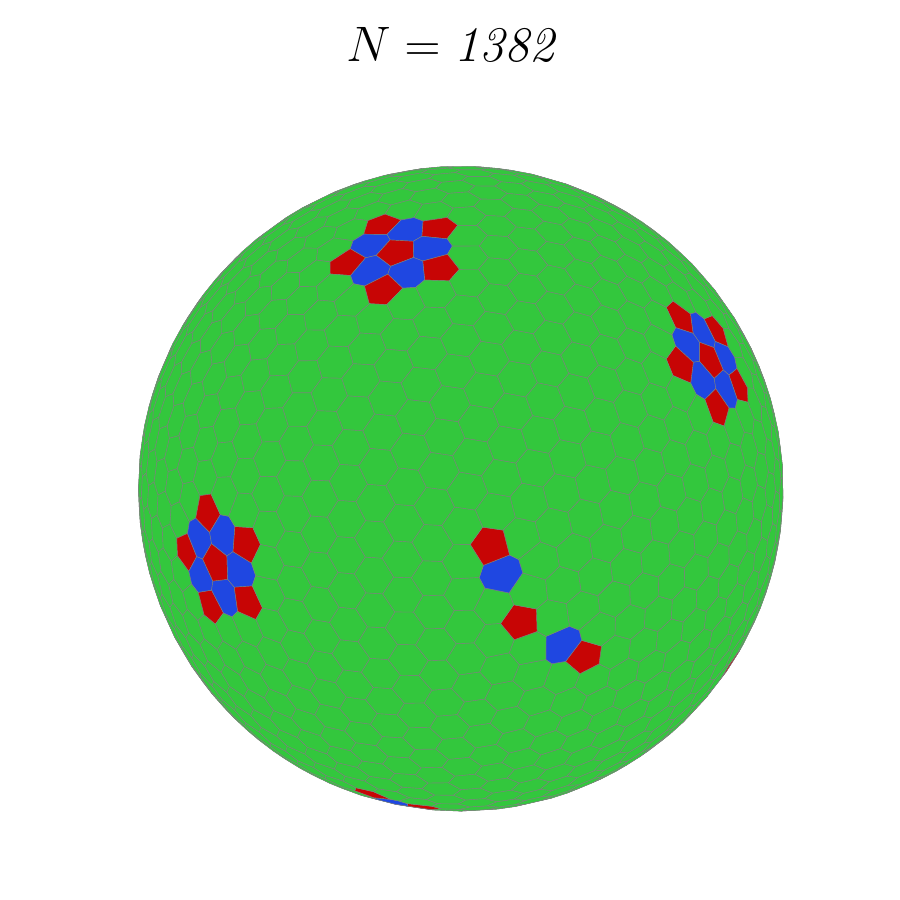}
\includegraphics[width=3.75cm]{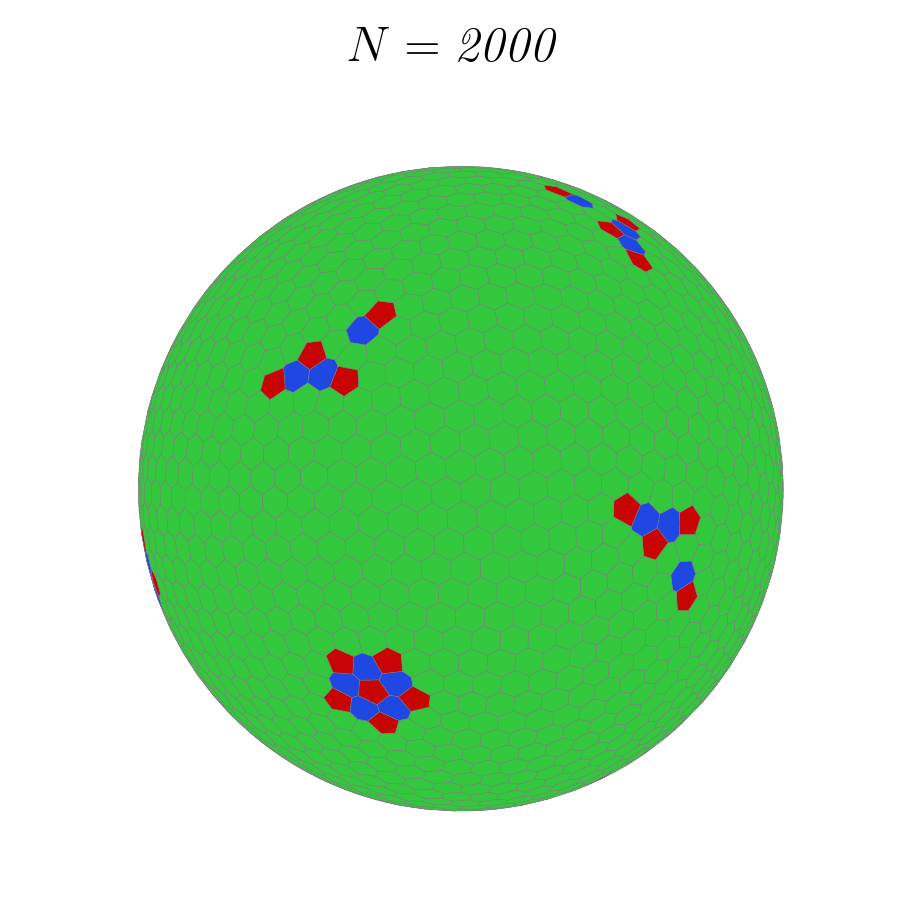}
\includegraphics[width=3.95cm]{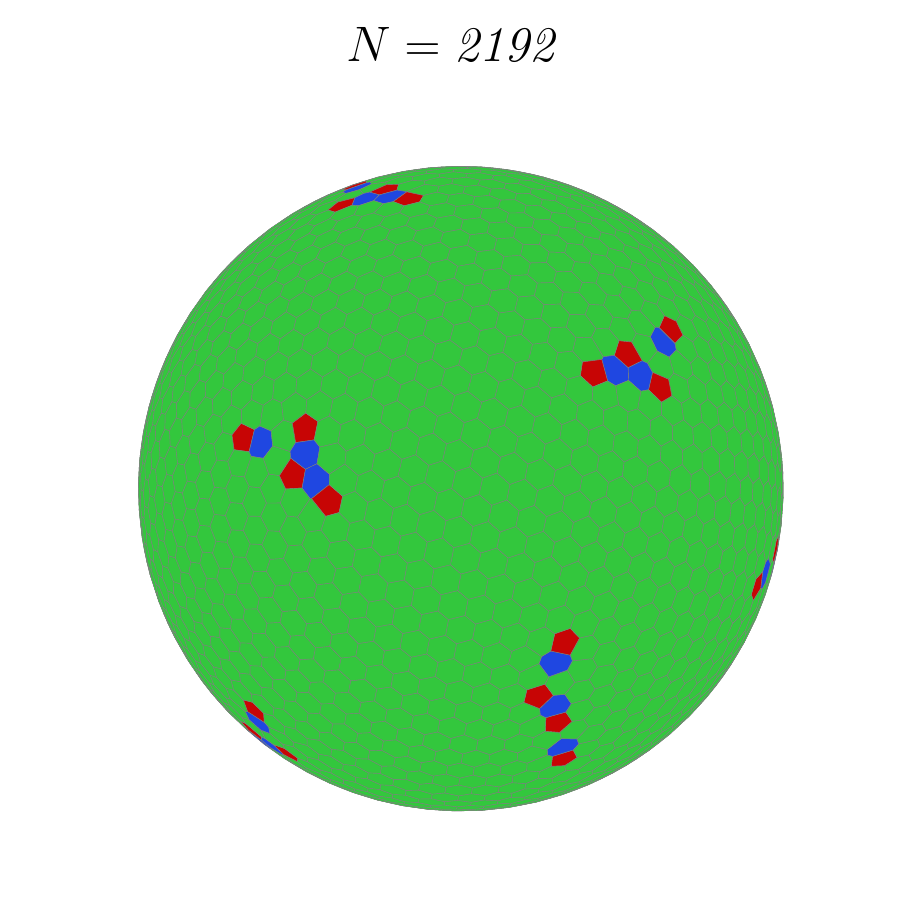}
\includegraphics[width=3.95cm]{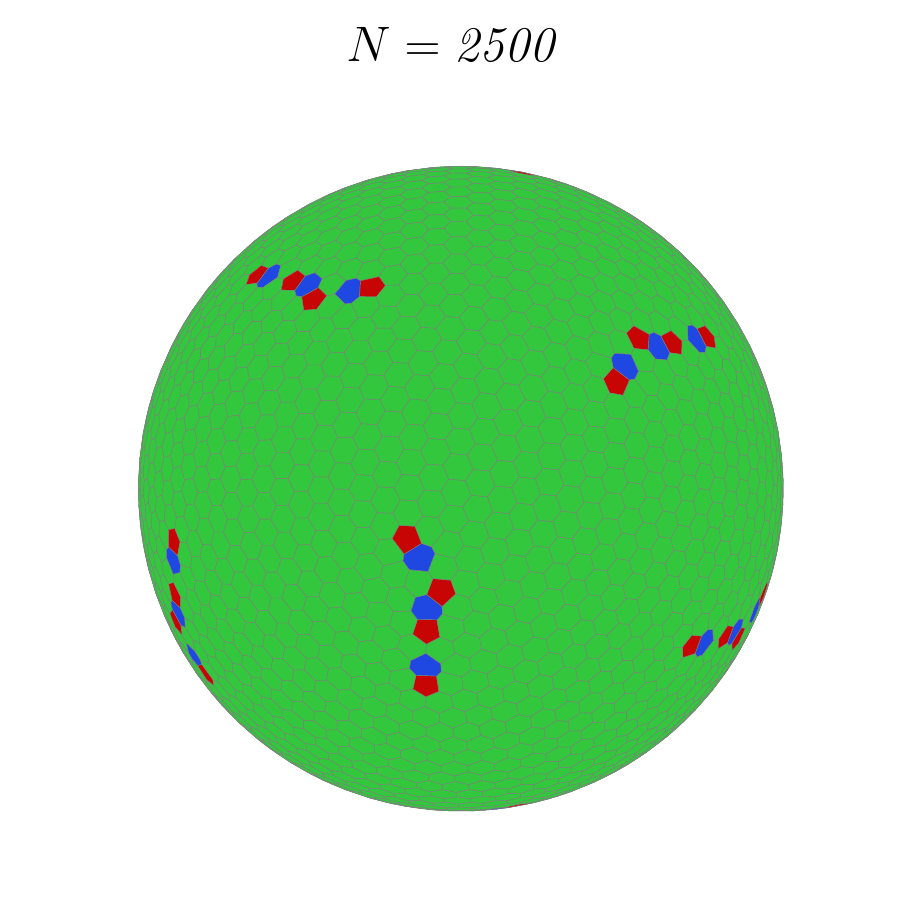}
\includegraphics[width=3.95cm]{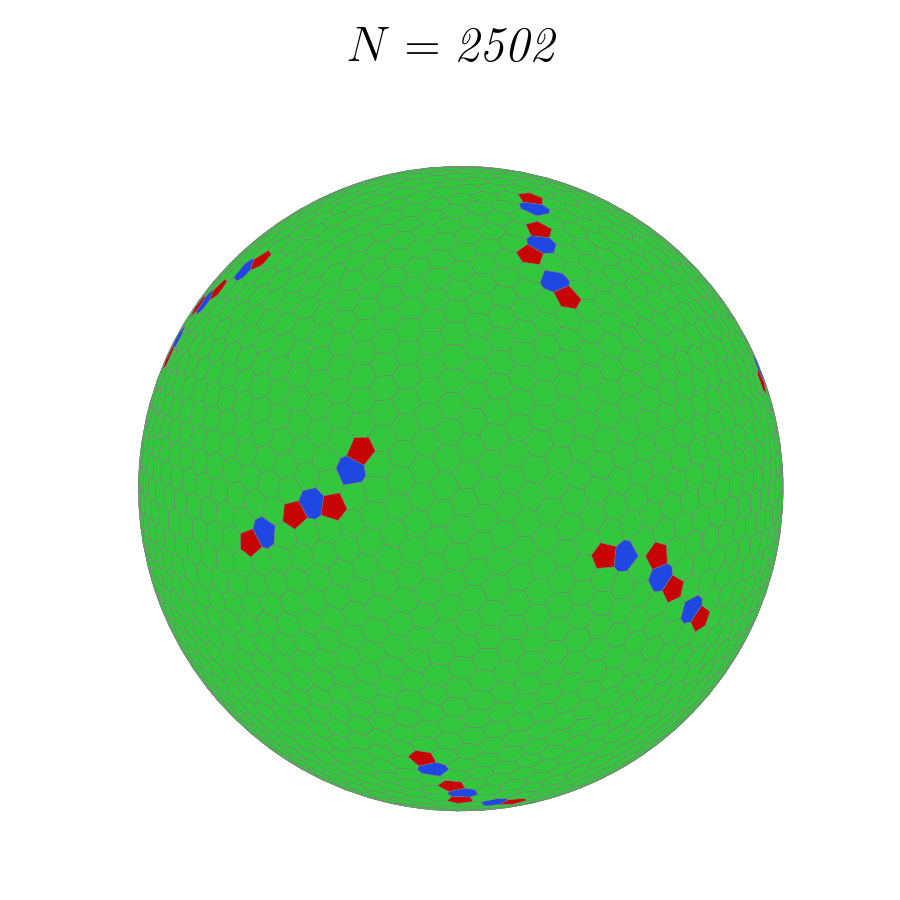}
\includegraphics[width=3.95cm]{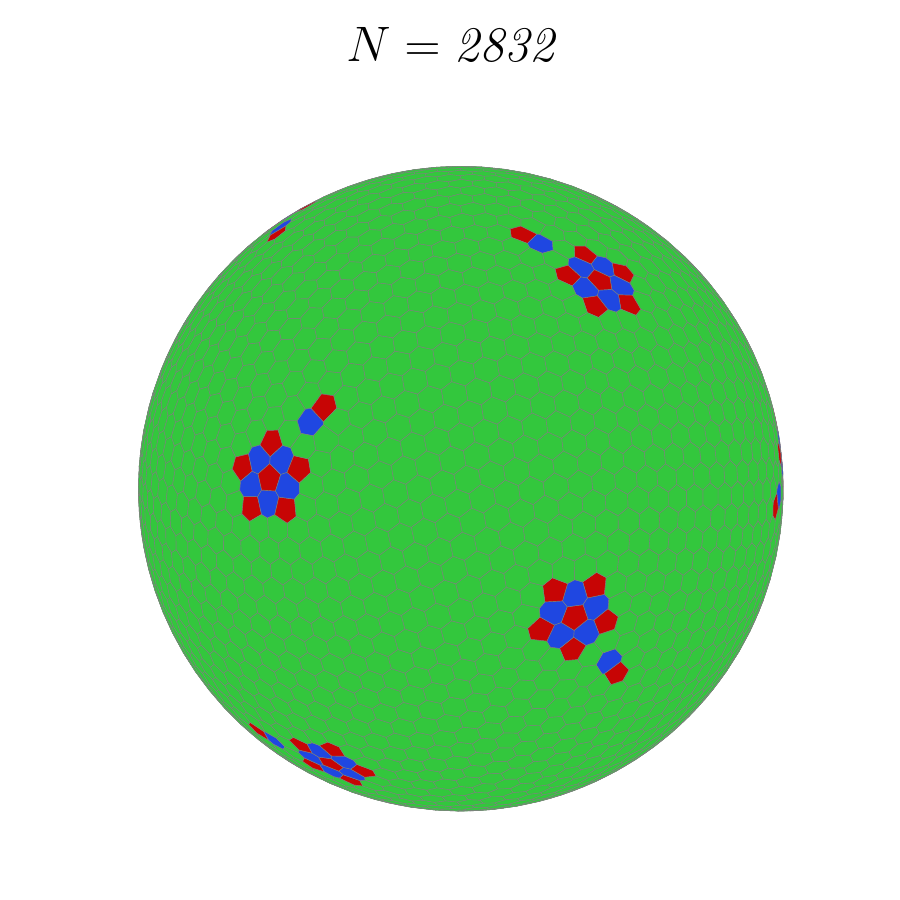}
\includegraphics[width=3.95cm]{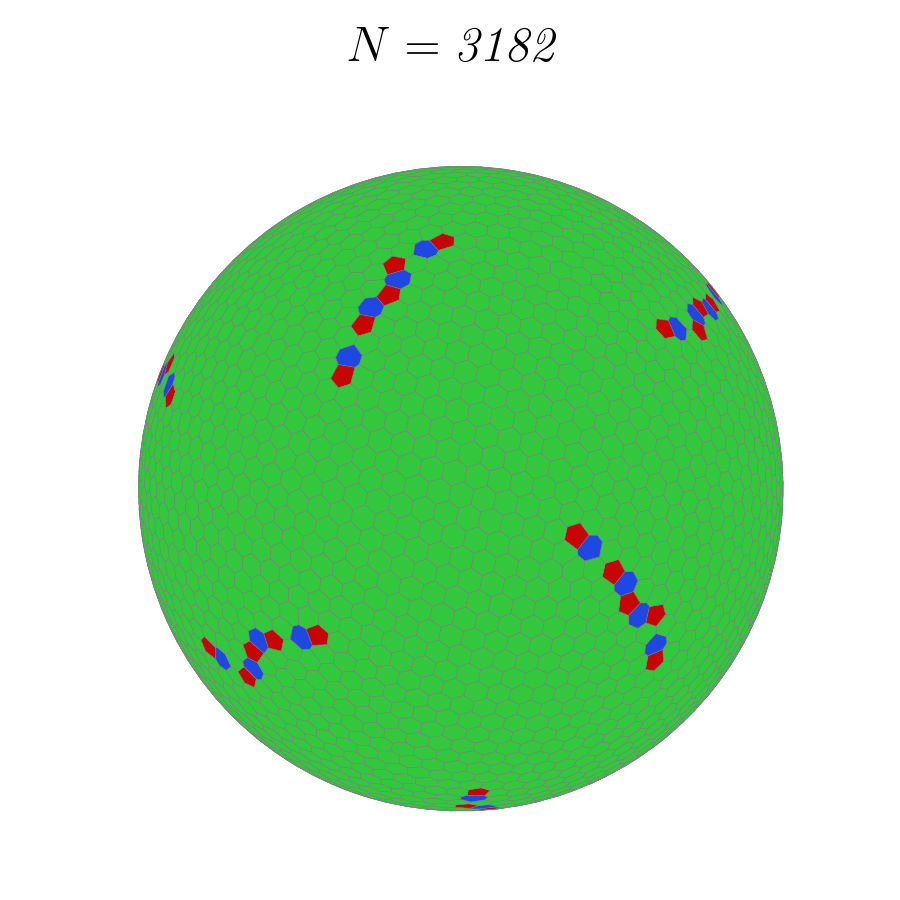}
\includegraphics[width=3.95cm]{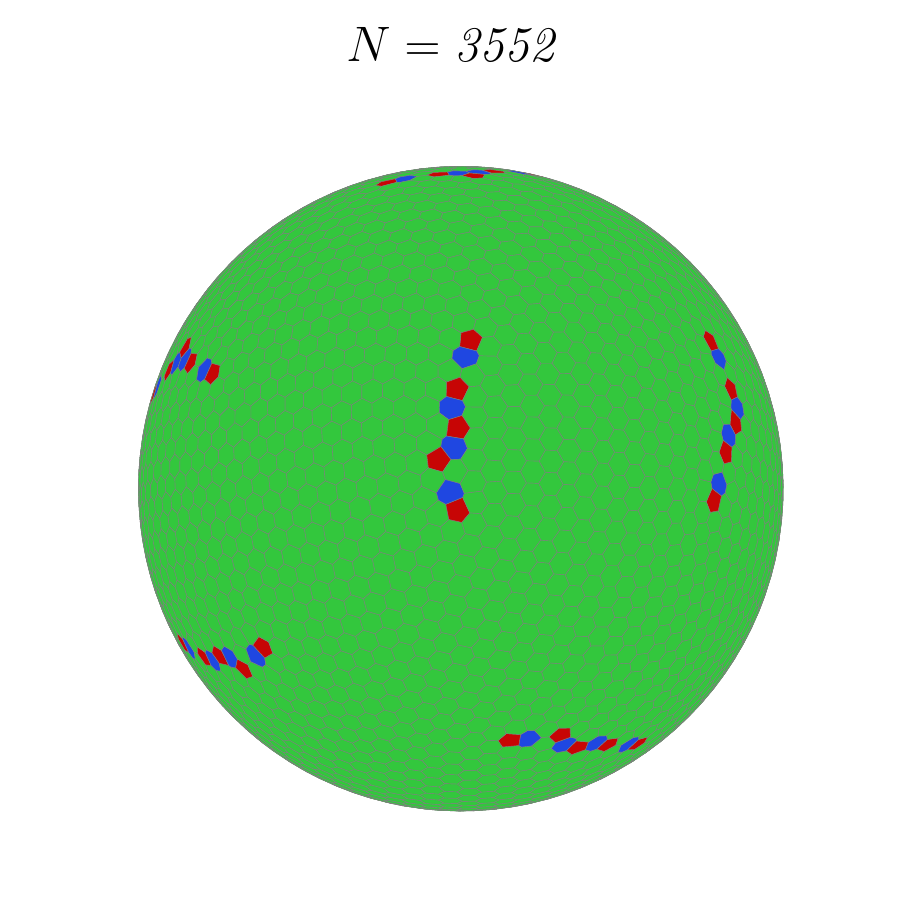}
\includegraphics[width=3.95cm]{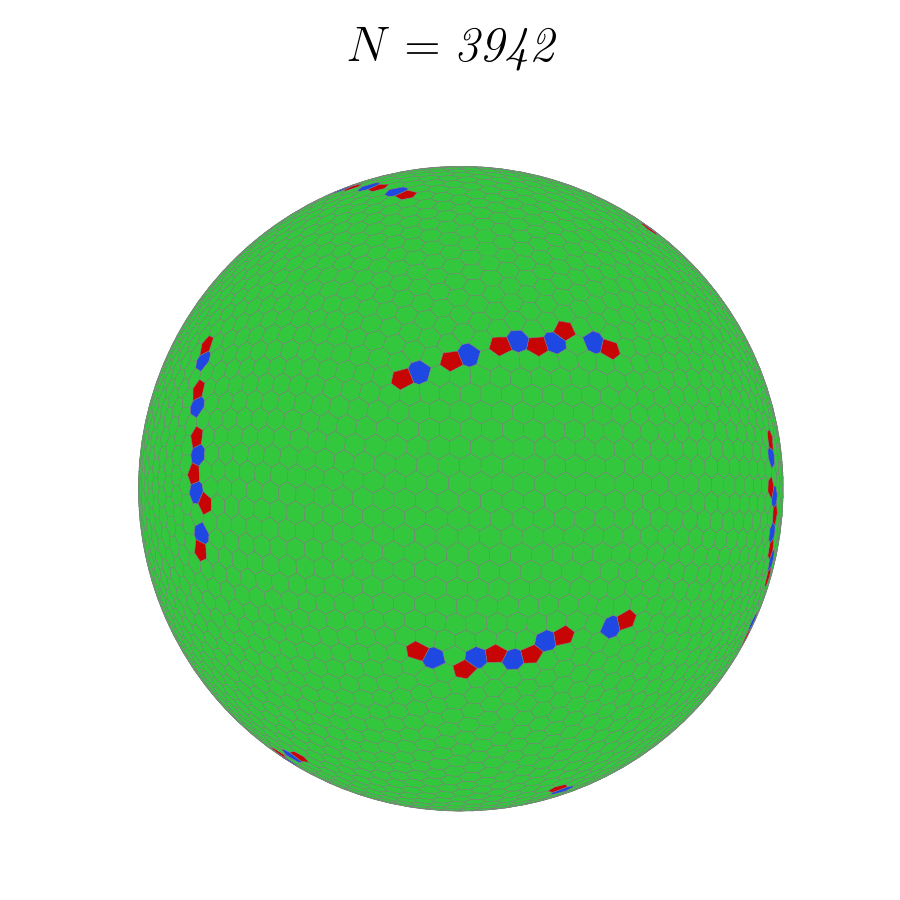}
\includegraphics[width=3.95cm]{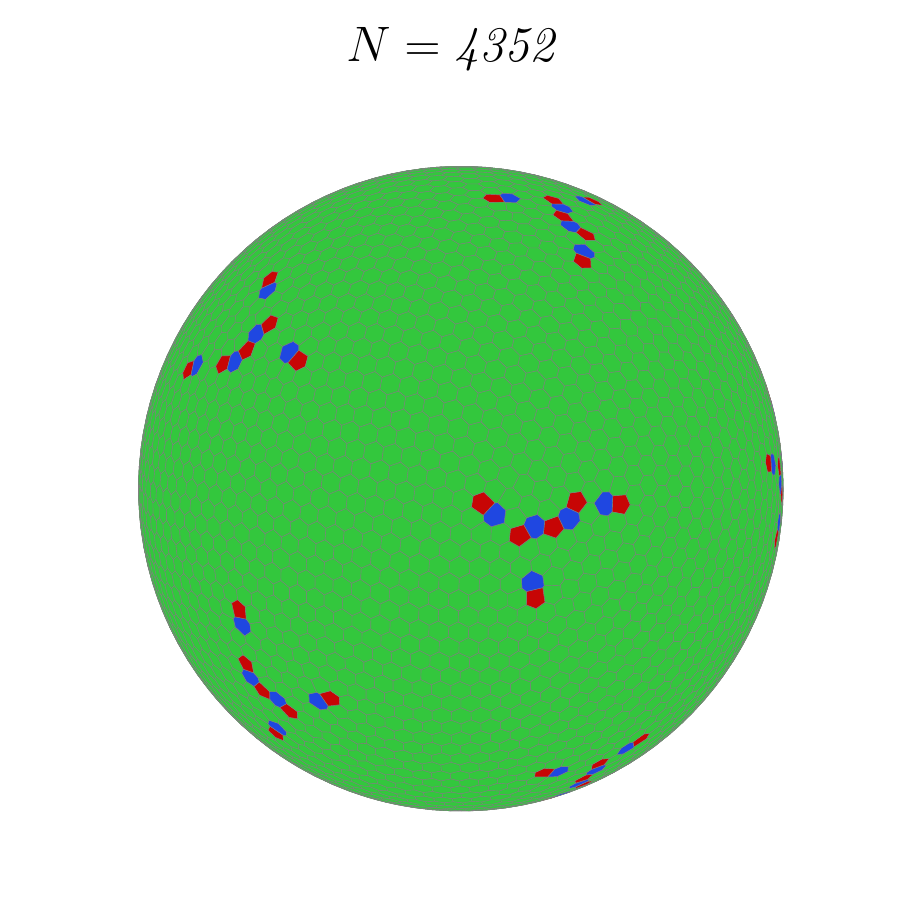}
\includegraphics[width=3.95cm]{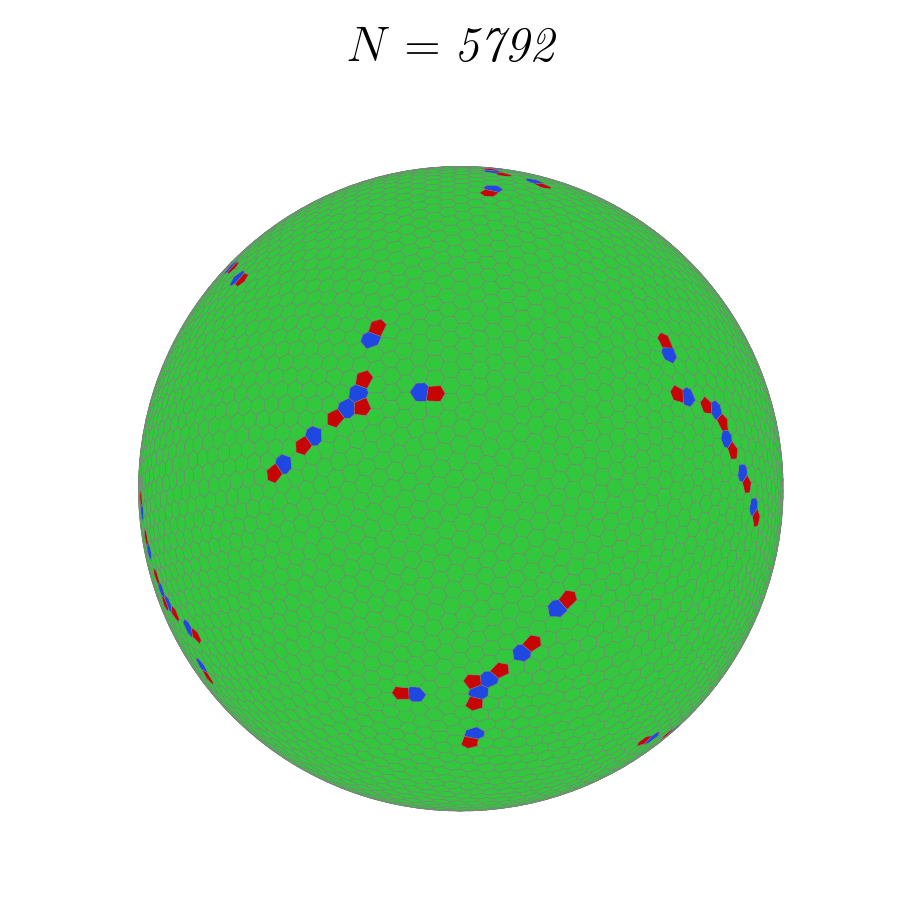}
\includegraphics[width=3.95cm]{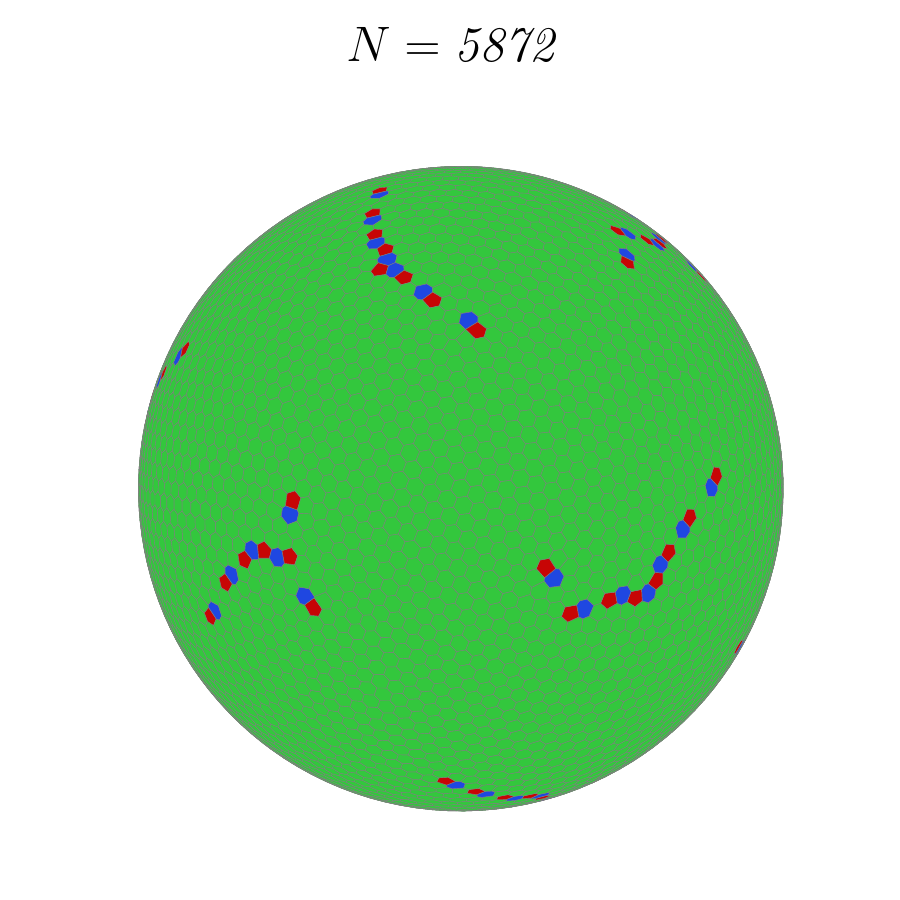}
\bigskip
\caption{Voronoi diagram for the lowest energy configurations obtained
for the Thomson problem at selected values of  $N$. }
\label{Fig_Voro_1}
\end{center}
\end{figure}

\begin{figure}
\begin{center}
\bigskip
\includegraphics[width=12.0cm]{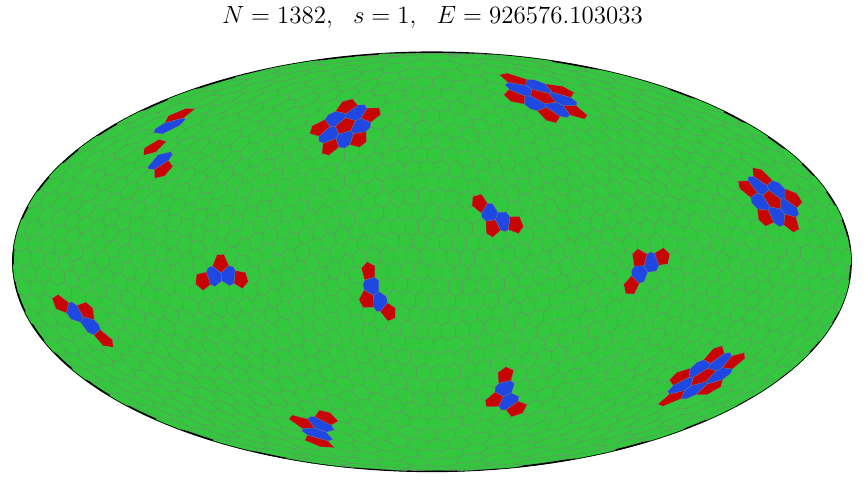} \\
\vspace{0.5cm}
\includegraphics[width=12.0cm]{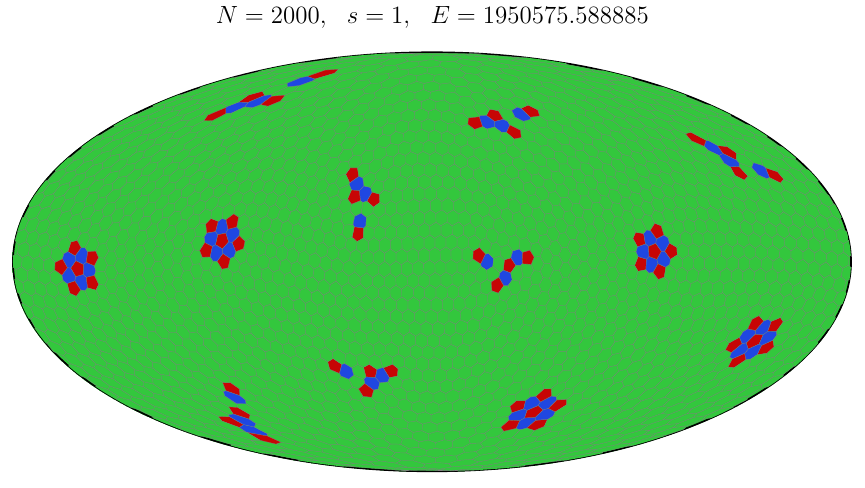} \\
\vspace{0.5cm}
\includegraphics[width=12.0cm]{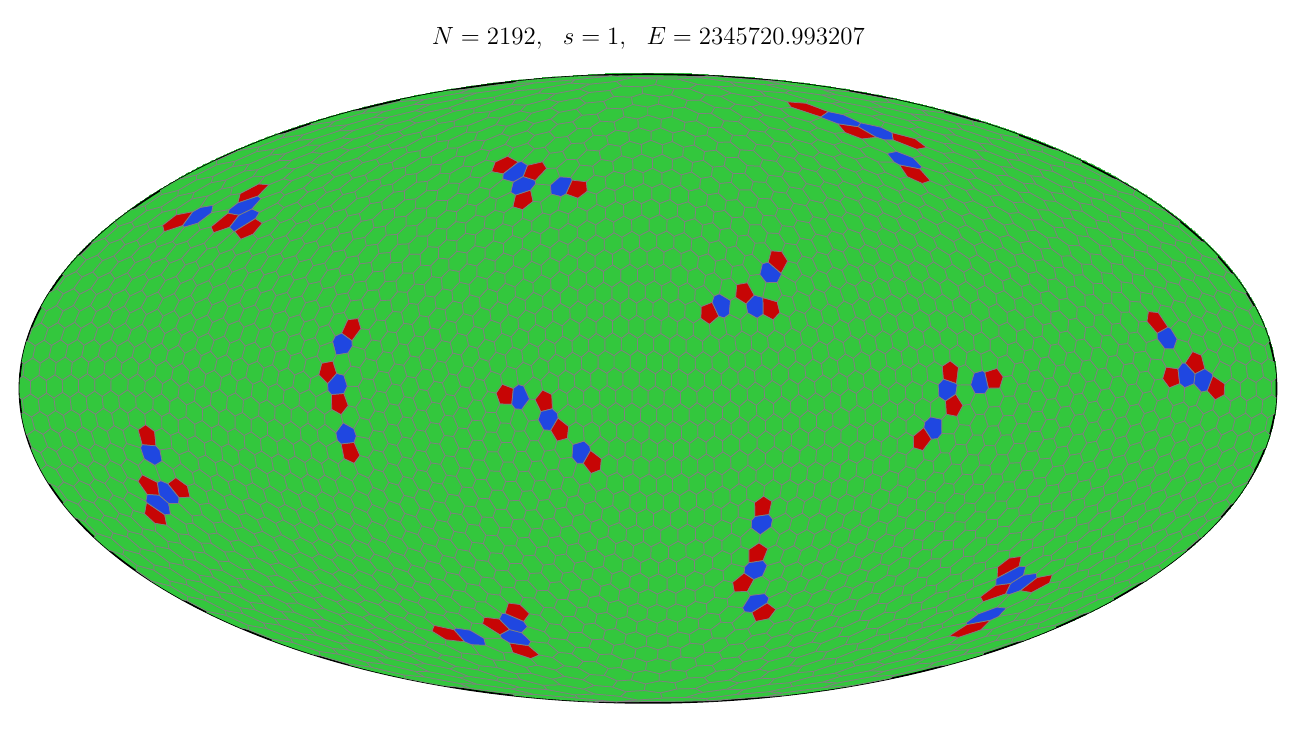} \hspace{0.2cm}
\bigskip
\caption{Mollweide projection of the Voronoi diagram for the lowest energy configurations
of the Thomson problem obtained in the present work with 1382, 2000, and 2192 charges.}
\label{Fig_Voro_1_mw_A}
\end{center}
\end{figure}

\begin{figure}
\begin{center}
\includegraphics[width=12.0cm]{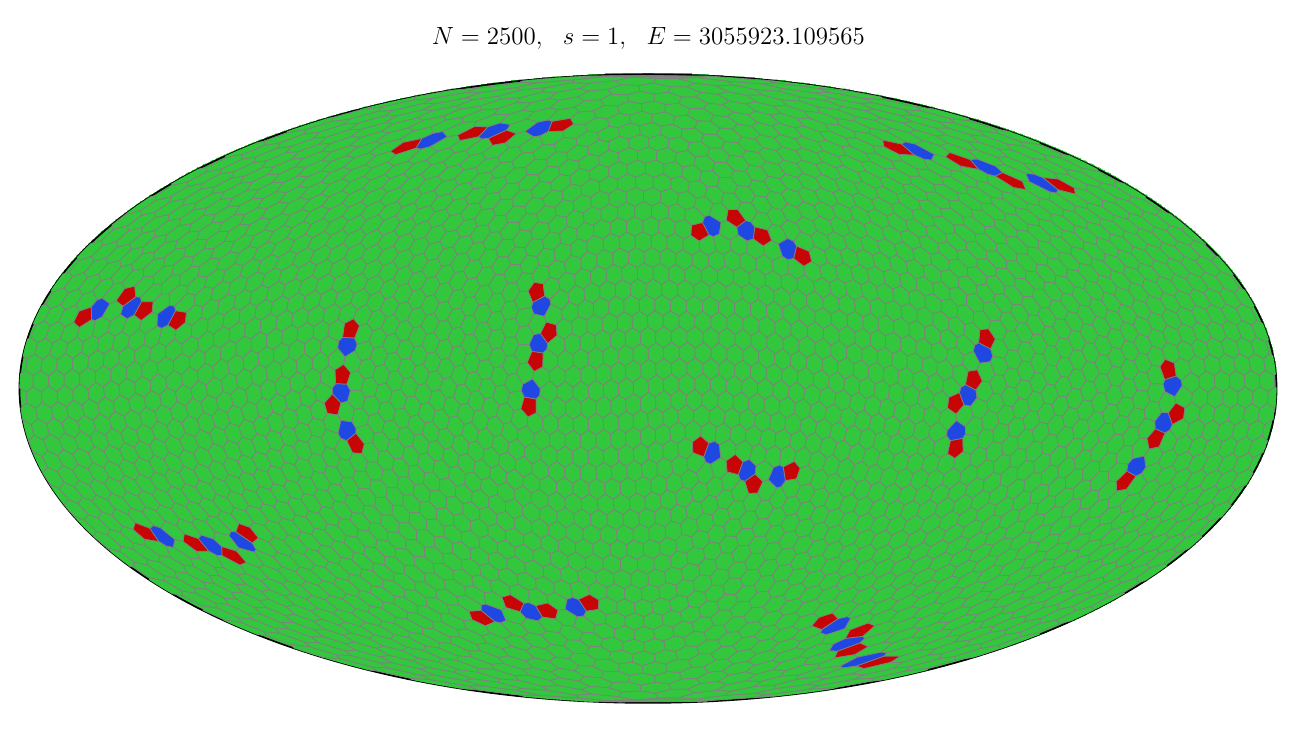} \\
\vspace{0.5cm}
\includegraphics[width=12.0cm]{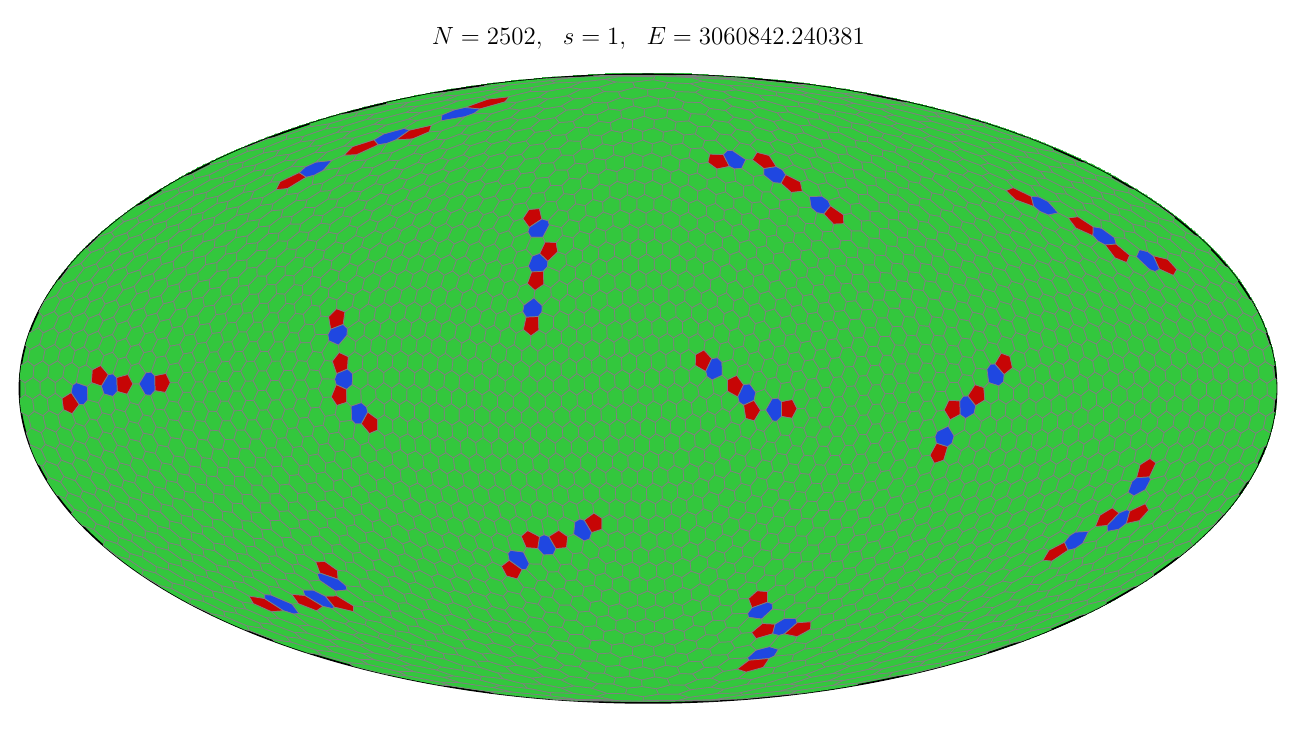} \\
\vspace{0.5cm}
\includegraphics[width=12.0cm]{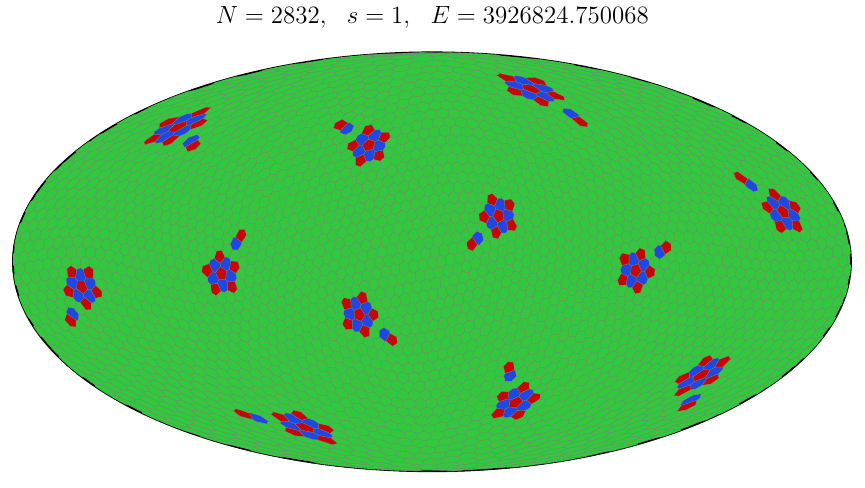} \\
\bigskip
\caption{Mollweide projection of the Voronoi diagram for the lowest energy configurations
of the Thomson problem obtained in the present work with 2500, 2502, and 2832 charges.}
\label{Fig_Voro_1_mw_B}
\end{center}
\end{figure}

\begin{figure}
\begin{center}
\includegraphics[width=12.0cm]{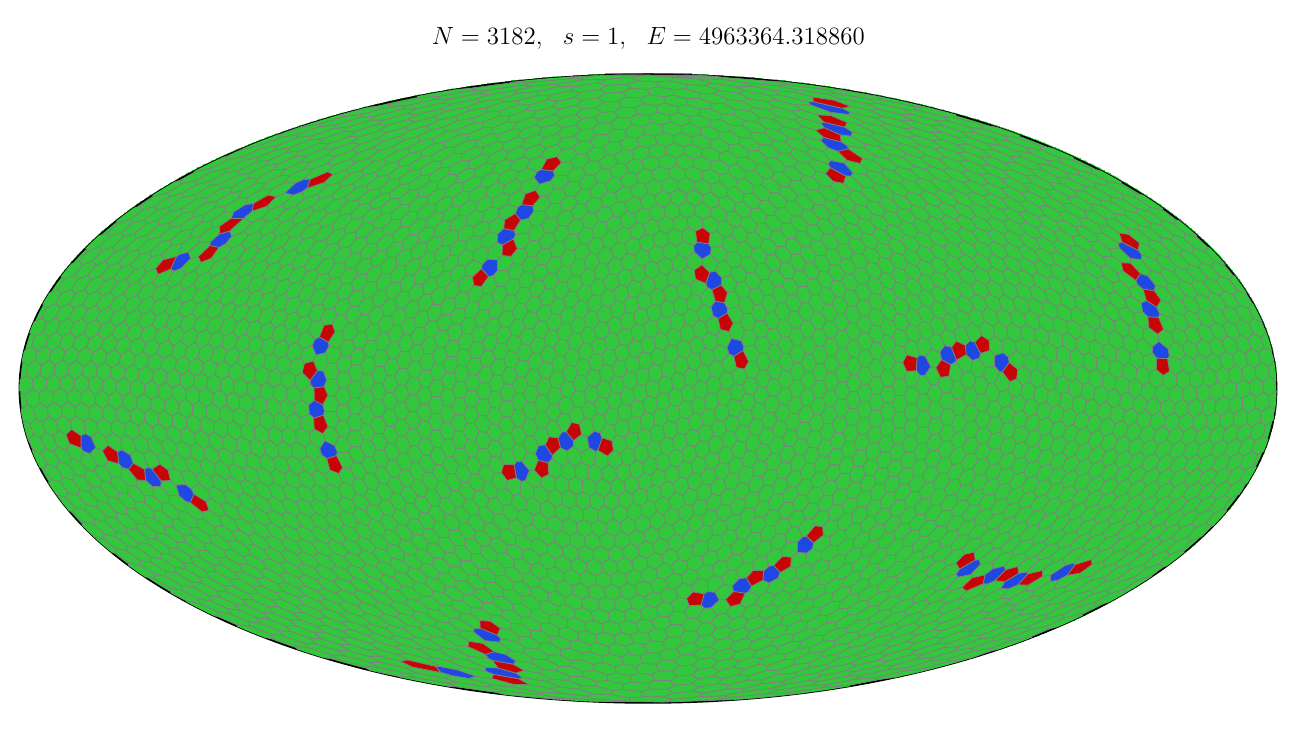} \\
\vspace{0.5cm}
\includegraphics[width=12.0cm]{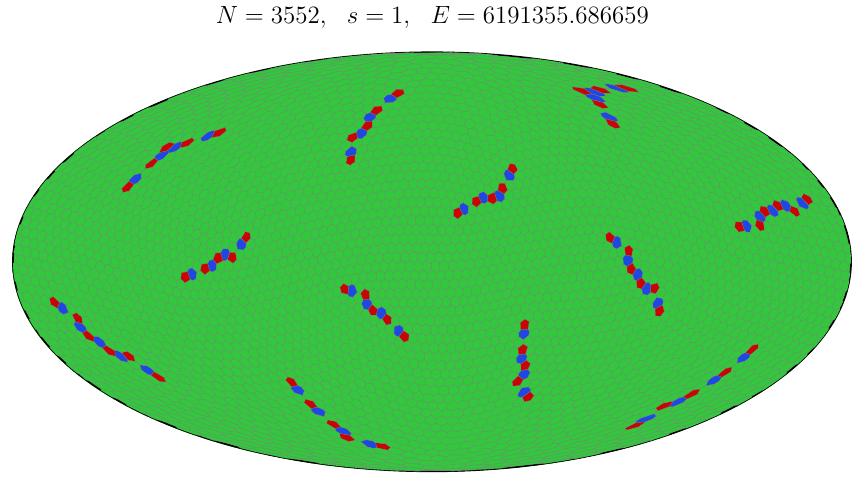} \\
\vspace{0.5cm}
\includegraphics[width=12.0cm]{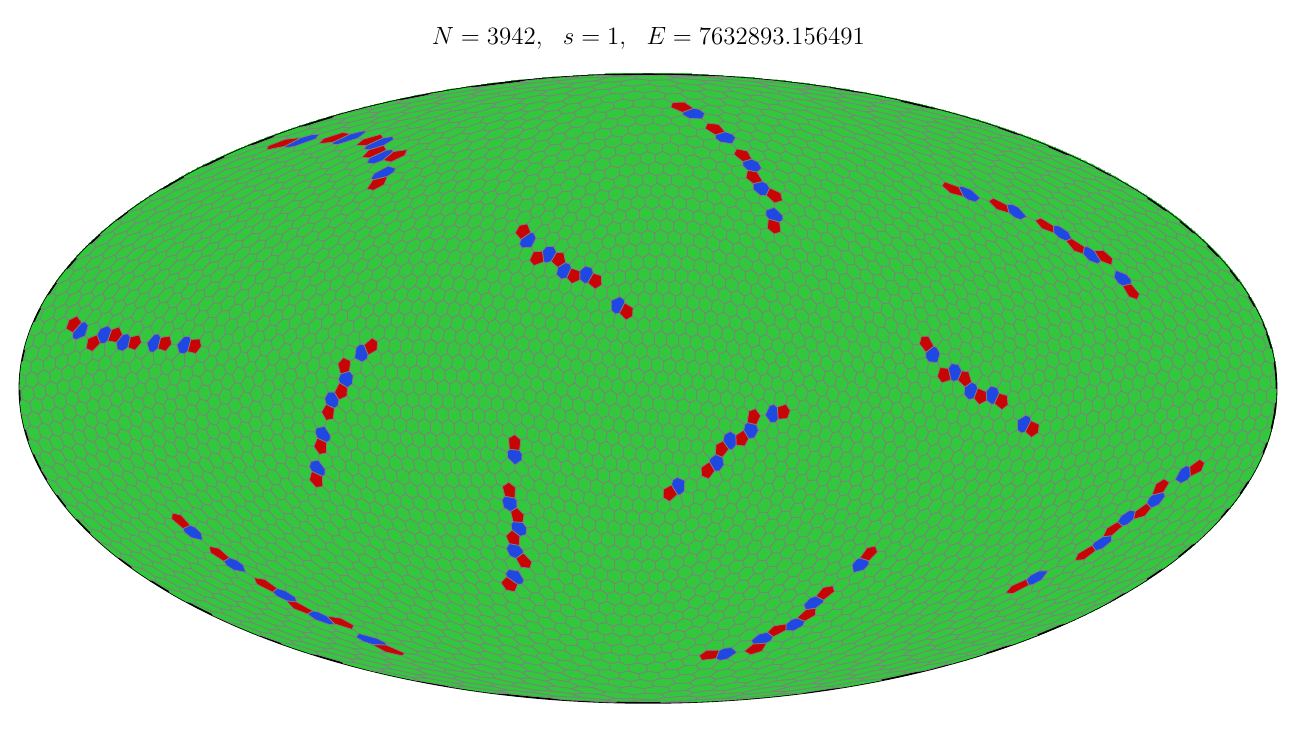} \\
\bigskip
\caption{Mollweide projection of the Voronoi diagram for the lowest energy configurations
of the Thomson problem obtained in the present work with 3182, 3552, and 3942 charges.}
\label{Fig_Voro_1_mw_C}
\end{center}
\end{figure}

\begin{figure}
\begin{center}
\bigskip
\includegraphics[width=12.0cm]{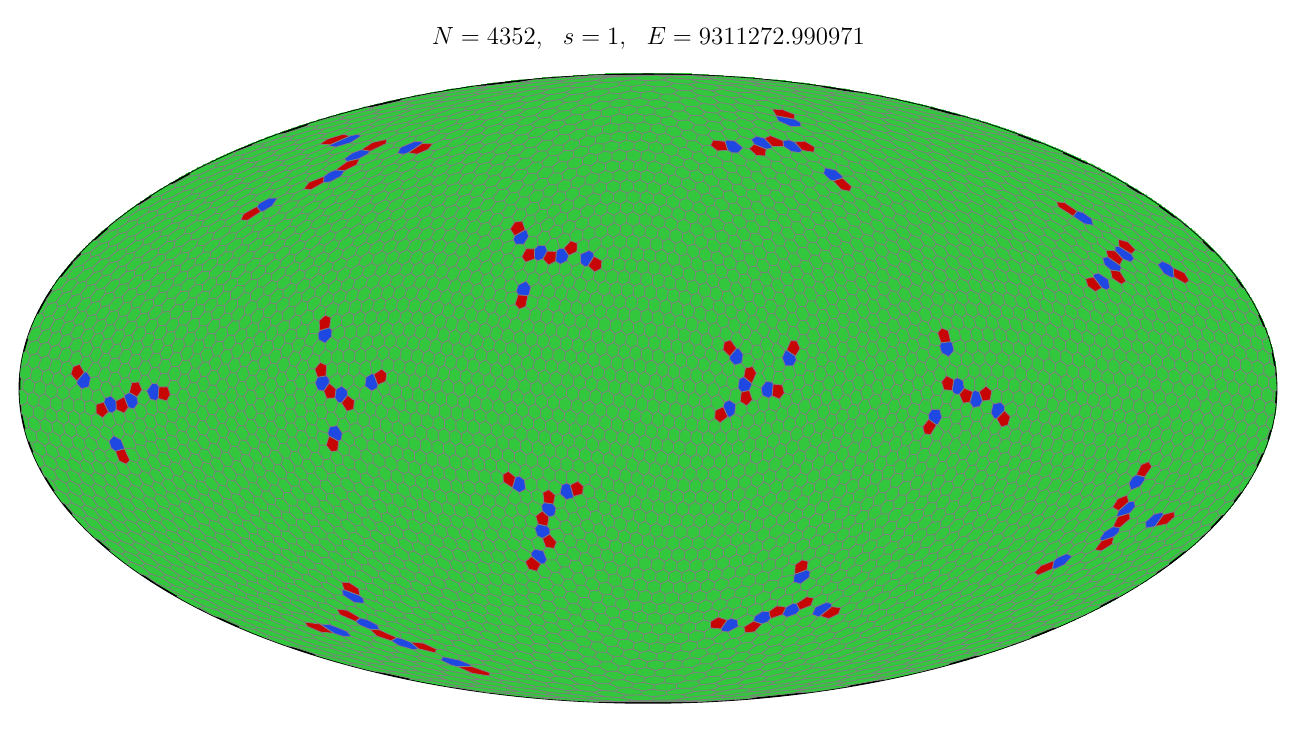} \\
\vspace{0.5cm}
\includegraphics[width=12.0cm]{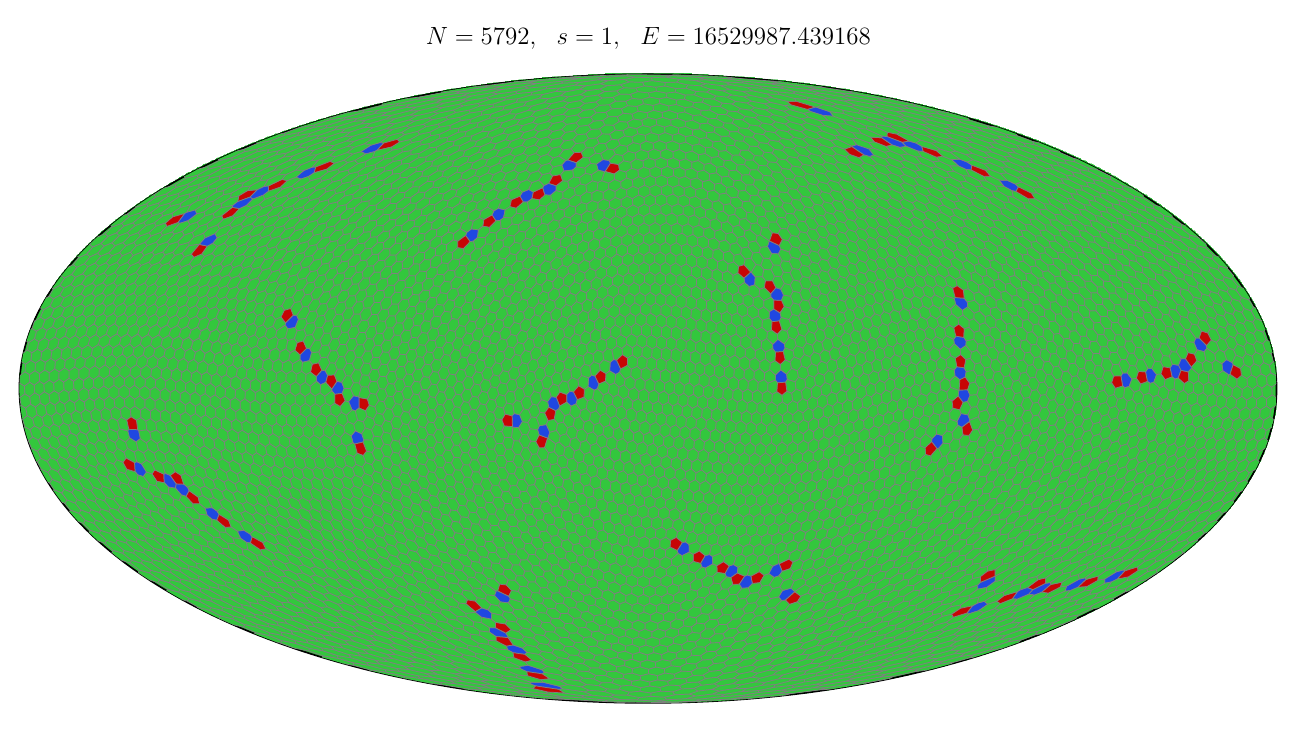} \\
\vspace{0.5cm}
\includegraphics[width=12.0cm]{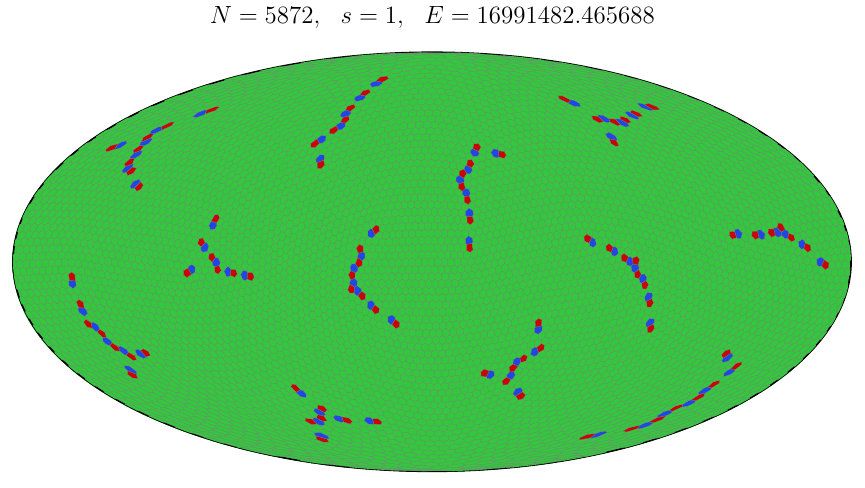} \\
\bigskip
\caption{Mollweide projection of the Voronoi diagram for the lowest energy configurations
of the Thomson problem obtained in the present work with  4352, 5792, and 5872.}
\label{Fig_Voro_1_mw_D}
\end{center}
\end{figure}

\begin{figure}
\begin{center}
\bigskip
\includegraphics[width=12.0cm]{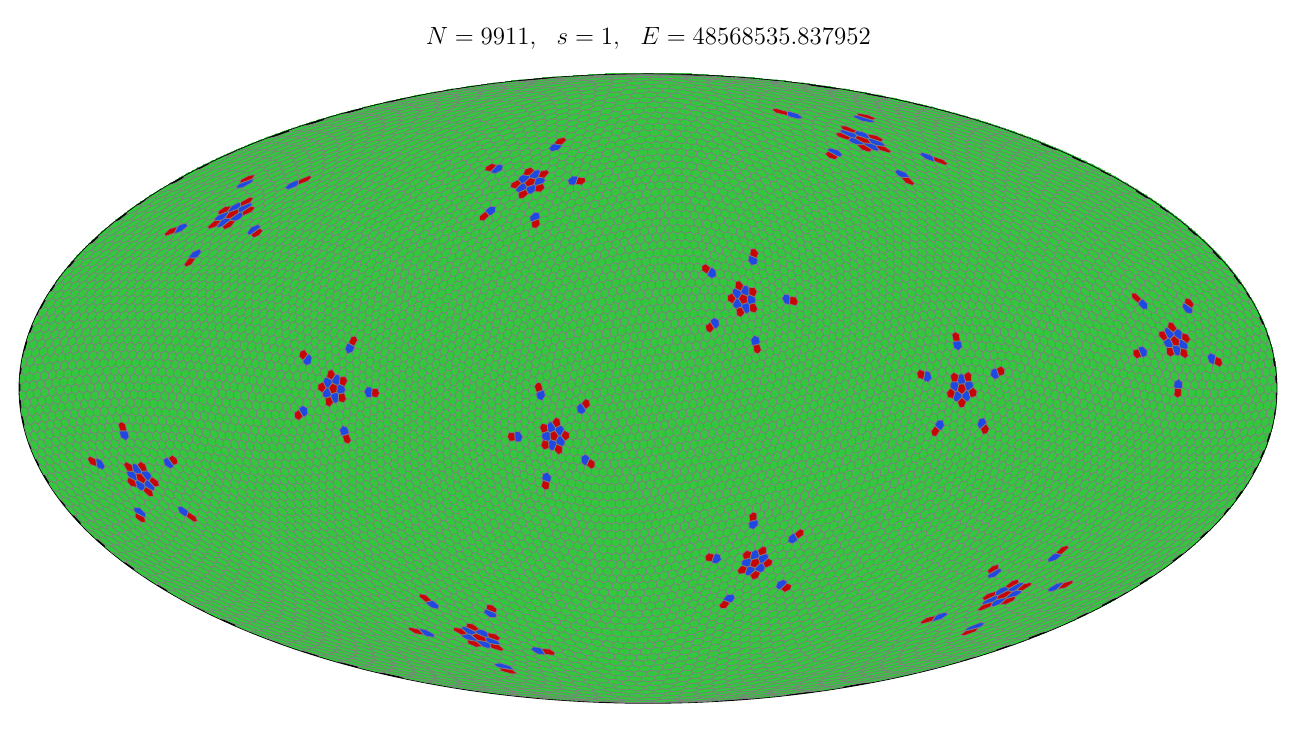} \\
\vspace{0.5cm}
\includegraphics[width=12.0cm]{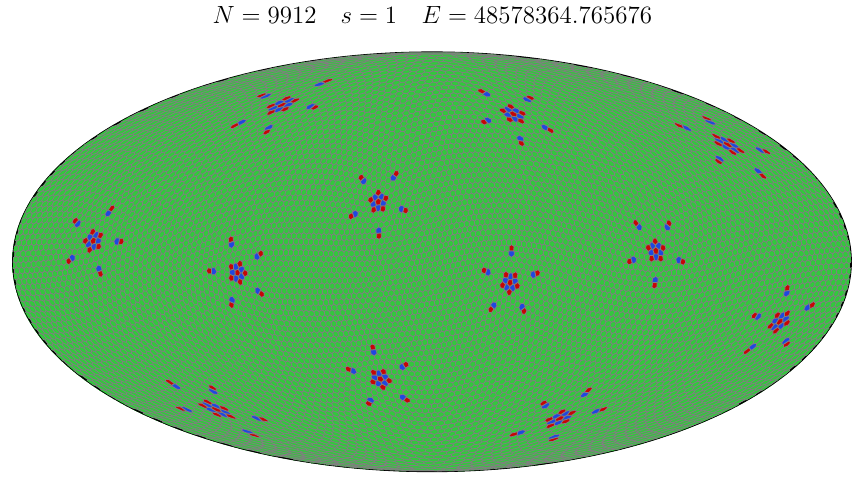} \\
\vspace{0.5cm}
\includegraphics[width=12.0cm]{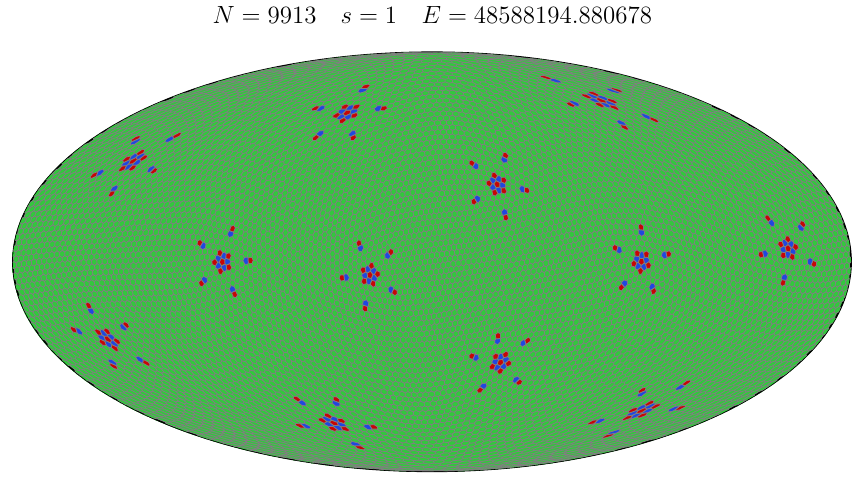} \\
\bigskip
\caption{Mollweide projection of the Voronoi diagram for the lowest energy configurations
of the Thomson problem obtained in the present work with  9911, 9912, and 9913.}
\label{Fig_Voro_1_mw_E}
\end{center}
\end{figure}

In Fig.~\ref{Fig_Voro_1} we plot the Voronoi diagrams of selected configurations that we have obtained
in the present work. 
Because one side of the sphere is hidden, these figures do not allow one to fully appreciate the complete defect structure. 
We have therefore used a Mollweide projection to represent the Voronoi
diagrams in a more efficient way (see
Figs.~\ref{Fig_Voro_1_mw_A},\ref{Fig_Voro_1_mw_B},\ref{Fig_Voro_1_mw_C}, \ref{Fig_Voro_1_mw_D} and \ref{Fig_Voro_1_mw_E}).

We have also considered two examples not previously studied in the literature, as far as we know.
In the first one, exploiting of a configuration of $5882$ points with rather low energy, we applied OSHBH over the region
$5863 \leq N \leq 5886$, managing to lower the energy substantially.
In the second test, we started with an $I$ symmetry configuration of $9912$ points and quite low energy and applied the algorithm over 
the range $9911 \leq N \leq 9920$. In this case, due to the larger system size 
we slightly modified the optimization step, moving to the
next $N$ as soon as a new lowest energy is found. This approach drastically reduces
the execution time, but it will also  lead to solutions of {\color{black} lower}  quality. 

\begin{figure}[!t]
\begin{center}
\includegraphics[width=10cm]{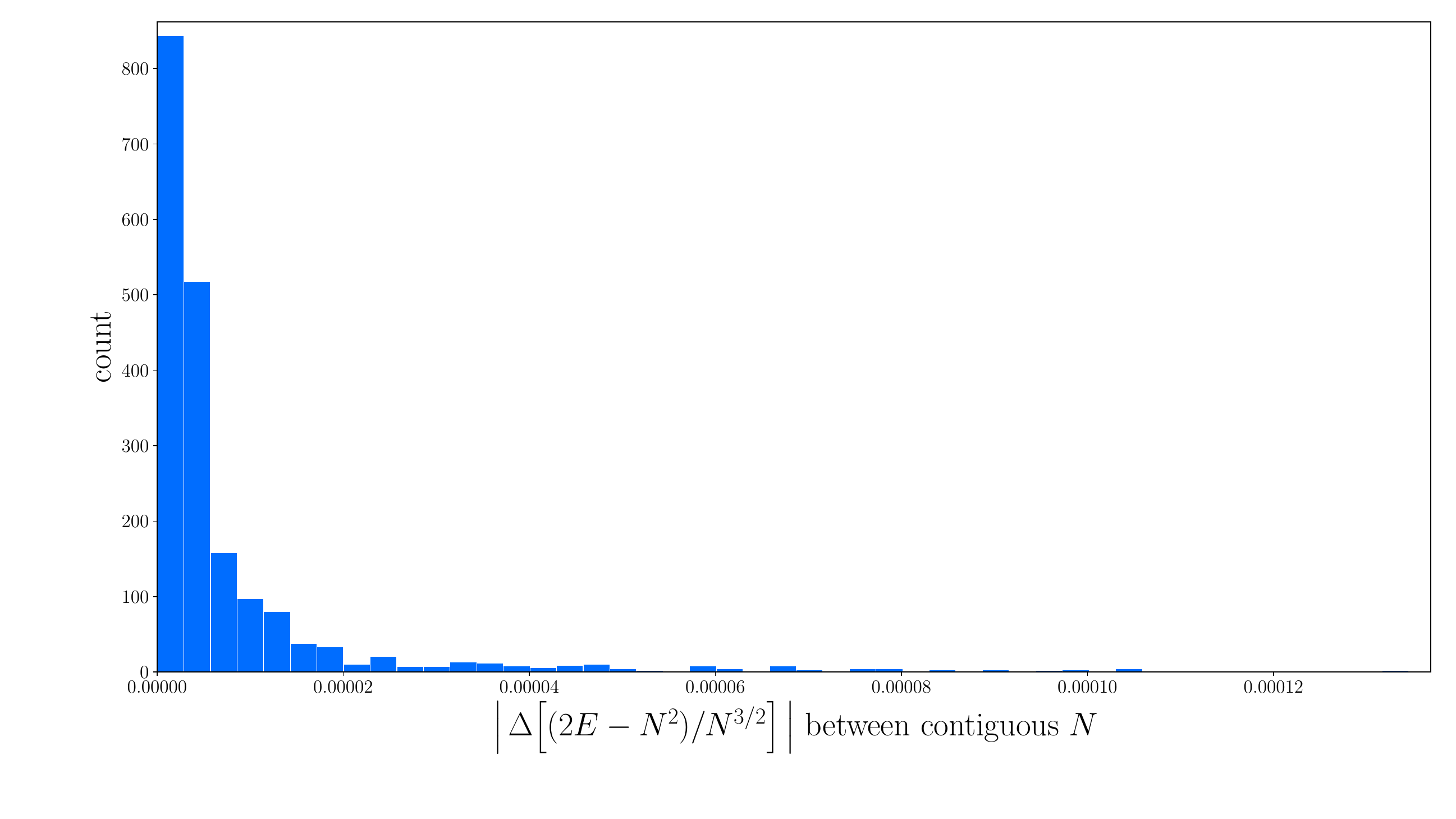}
\caption{Difference in scaled energy $\mathcal{Y}_N  = (2 E_N-N^2)/N^{3/2}$ between contiguous $N$ for $N \leq 2000$.}
\label{Fig_deltaY}
\end{center}
\end{figure}

In Fig.~\ref{Fig_deltaY} we plot the difference 
\[
\Delta \mathcal{Y}_N  = | \mathcal{Y}_{N+1} -\mathcal{Y}_N|
\]
where $N \leq 2000$ and $\mathcal{Y}_N \equiv (2 E_N-N^2)/N^{3/2}$ is the scaled energy
calculated using the set of configurations  that we have calculated (not 
necessarily global minima but expected to have low energy). 
The point we want to make here is that one expects $\Delta \mathcal{Y}$ to be rather small for the
lowest energy configurations with continuous $N$. This result implies that, if one has
found a configuration of $N$ points with particularly low energy, it is quite
likely that the configurations with neighbouring $N$ will have a similar
(scaled) energy. The fact that the algorithm manages to find solutions with 
quite low energy while bouncing over a range of values for $N$ is the empirical confirmation of this intuition.

\begin{figure}
\begin{center}
\includegraphics[width=11cm]{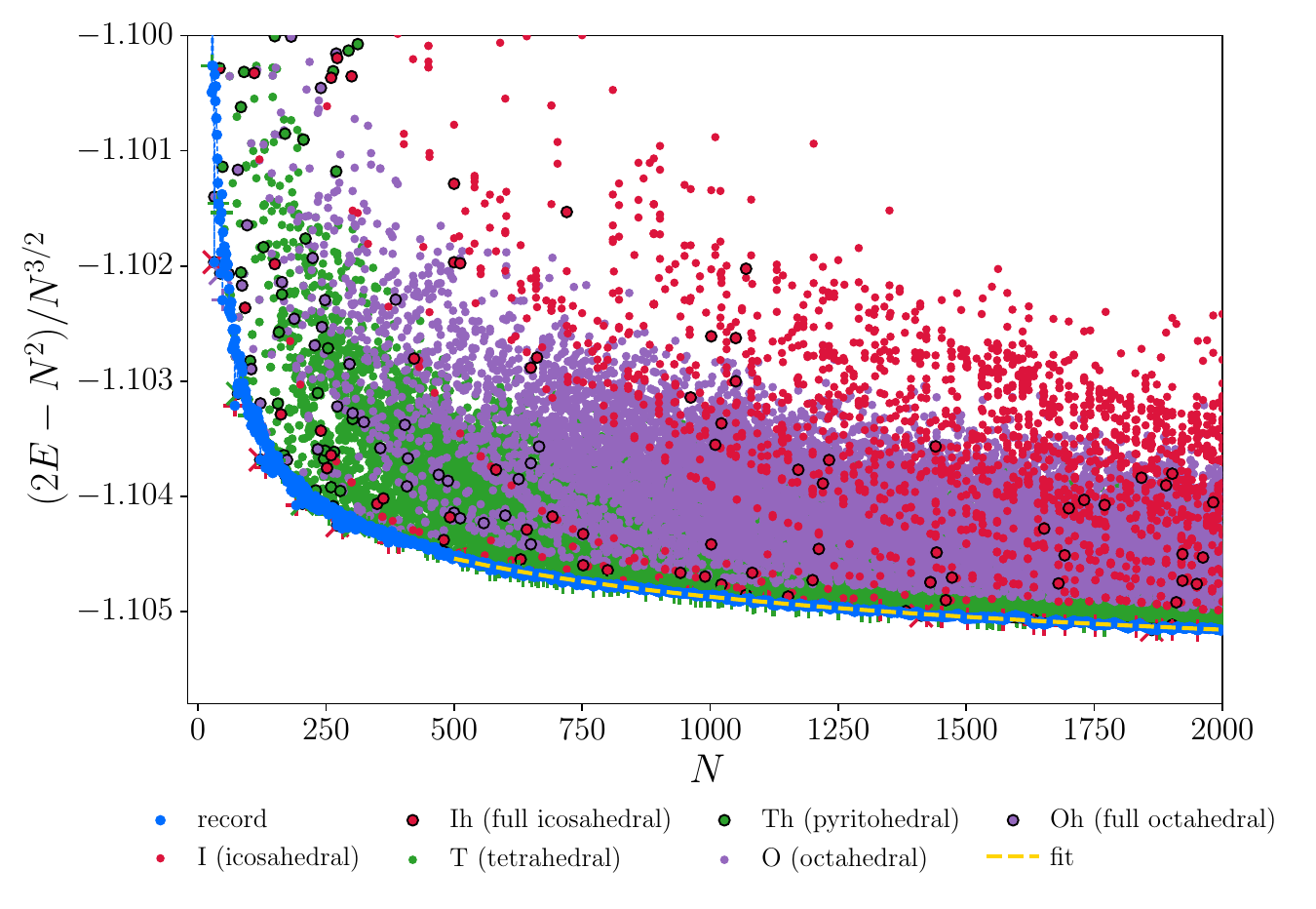}
\caption{$\mathcal{Y}_N = (2 E_N-N^2)/N^{3/2}$ as a function of $N$.}
\label{Fig_asymptotics}
\end{center}
\end{figure}

In Fig.~\ref{Fig_asymptotics} we plot $\mathcal{Y}_N$  for values $N \leq 2000$. The blue points in this plot correspond
to the best results obtained running our algorithm, while points of different colors correspond to configurations of different 
symmetry (lowest energy points marked with crosses correspond to symmetric
configurations). Although we have not carried out systematic global
optimization for all these configurations, 
the blue points provide a useful upper limit: in particular, the energy is expected to behave  for large $N$ as
\[
E_N \approx \left(  \frac{N^2}{2}- \mathcal{C} N^{3/2} +\dots \right) \ .
\]

The conjectured value of $\mathcal{C}$~\cite{Saff97} is
\begin{equation}
\begin{split}
\mathcal{C}  &\equiv - 3 \left( \frac{\sqrt{3}}{8\pi}\right)^{1/2}
\zeta(1/2) \sum_{n=0}^\infty \left( \frac{1}{\sqrt{3n+1}} - \frac{1}{\sqrt{3n+2}}\right) \\
&= \frac{3^{3/4} \zeta \left(\frac{1}{2}\right) \left(\zeta
   \left(\frac{1}{2},\frac{2}{3}\right)-\zeta
   \left(\frac{1}{2},\frac{1}{3}\right)\right)}{2 \sqrt{2 \pi }} \\
&\approx 0.5530512934
\end{split}
\end{equation}
and $\zeta(s,a)$ is the generalized Riemann zeta function. 

The dashed yellow curve is the least square fit of the blue points for $500 \leq N \leq 2000$:
\begin{equation}
f(N) = -1.1060 + \frac{0.0443}{\sqrt{N}} - \frac{0.2488}{N}
\label{fit}
\end{equation}
where the constant term is very close to $2 \mathcal{C} = 1.1061025868$.

\begin{figure}
\begin{center}
\includegraphics[width=11cm]{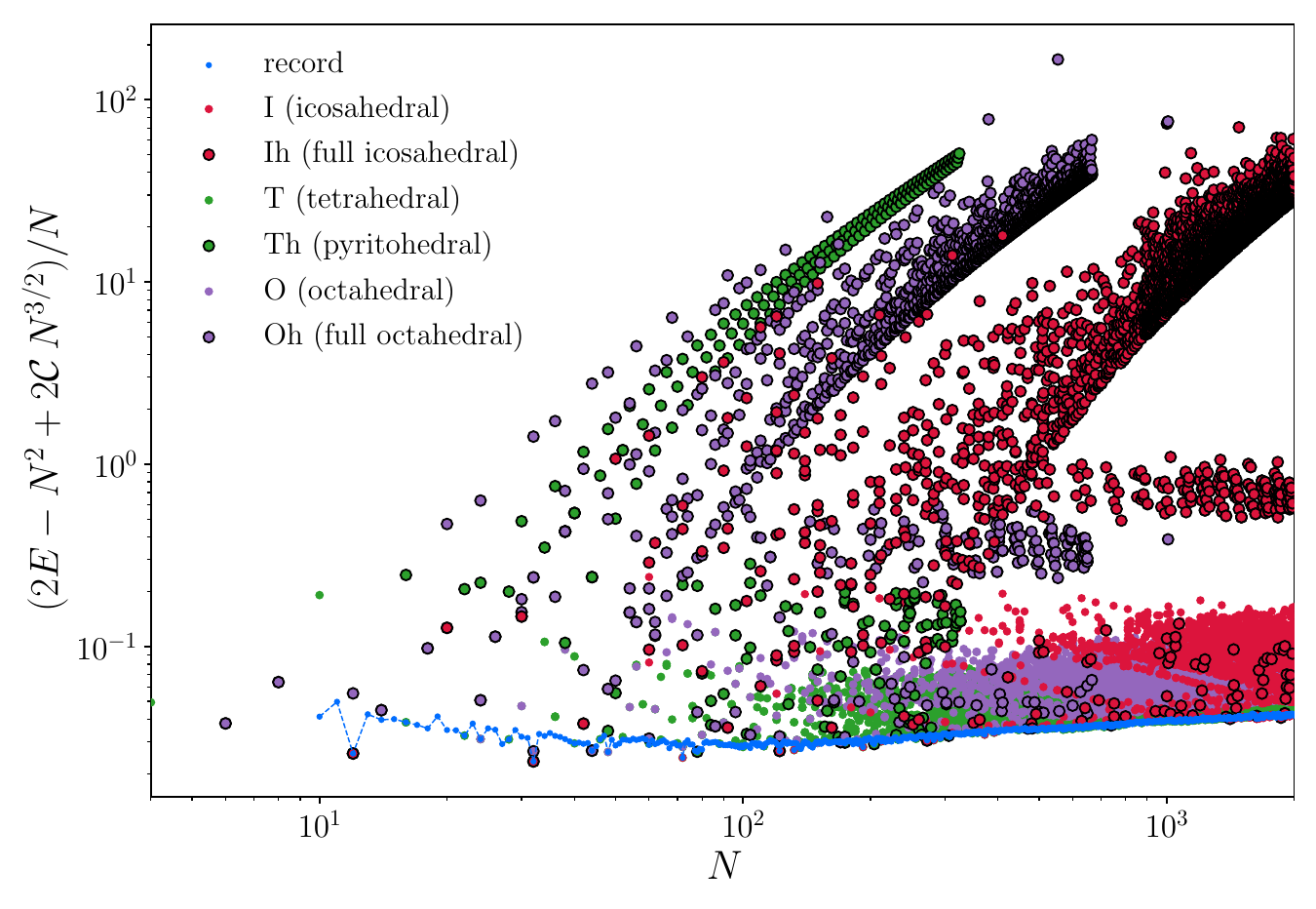}
\caption{$(2 E_N-N^2+2 \mathcal{C} N^{3/2})/N$ as a function of $N$.}
\label{Fig_conjecture}
\end{center}
\end{figure}

In Fig.~\ref{Fig_conjecture}  we plot the quantity $(2 E_N-N^2+2 \mathcal{C}
N^{3/2})/N$ for the configurations that we have obtained for $N \leq 2000$. In
this case we can see the presence of different behaviours, particularly
associated with solutions for point group 
symmetries $I_h$, $O_h$ and $T_h$. Solutions with symmetries $I$, $O$ and $T$,
on the other hand, are found to belong to the lowest group. We
expect each group to evolve to a different limit for $N \rightarrow \infty$,
and the competition between them in the size range considered here is already intriguing.

\section{Conclusions}
\label{sec_concl}

We have presented a computationally inexpensive algorithm designed to locate
low energy local minima of the Thomson problem for thousands of particles.
For smaller sizes where previous basin-hopping\cite{lis87,walesd97a,waless99} global optimisation results exist, 
we can benchmark the current approach, which reliably produces competitive structures.
Our approach sweeps up and down through selected size ranges, strategically adding or removing a charge.
New lower energy solutions then propagate through the size range, and
can be seeded with known structures or local minima from high symmetry point groups
that can be constructed systematically at particular sizes.
We find that exploiting high symmetry seems to be a particularly effective strategy, which provides another
illustration of the principle of maximum symmetry.\cite{wales98,wales98b,OakleyJW13}

Since the current approach steps between local minima we can think of it as a variety of
basin-hopping.\cite{lis87,walesd97a,waless99} 
In the step-taking we propose moves that exploit information from neighbouring sizes, and
high symmetry seeds. The changes in size also provide the opportunity to transplant favourable
defect motifs, which bears some resemblance to the cut-and-splice moves used to generate
offspring in previous genetic algorithm global optimisation approaches.\cite{deavenh95,deaventmh96}
These Lamarckian genetic algorithms\cite{MorrisGHHHBO98, Johnston03} considered a population of local minima, 
and so they correspond to basin-hopping with evolutionary moves.
Overall, the framework we have used in the present work corresponds to an
Oscillating Size Hybrid Basin-Hopping (OSHBH) approach, which retains the essential basin-hopping 
foundation of steps between local minima.\cite{waless99}
The propagation of information between sizes can also occur in grand canonical basin-hopping,\cite{CalvoSW16}
but in that approach steps between parallel replicas depend on a chemical potential.

In the larger size regime OSHBH has significantly improved upon previous solutions.
However, as the runs proceed we are routinely finding lower minima at some sizes. 
We also envisage improvements in the future based on systematic basin-hopping with conventional
random perturbations starting from the best known solutions, and perhaps in more sophisticated moves
that further exploit symmetry and transplantation of the defects.
Nevertheless, the current results already suggest new defect packing schemes that may find applications
in a variety of systems with spherical topology.
We are therefore reporting the best solutions at the current point in time, and we will
provide a continuous update of lowest minima in the Cambridge Landscape Database,
where they will be available for open access download.\cite{CLD}
Additional resources, including icosahedral nets for constructing models, and all the
code we have used, will also be made available at the same web site.

\section*{Data availability}
The lowest minima that we obtain for the Thomson problem will be systematically updated 
in the Cambridge Landscape Database.\cite{CLD}


\section*{Acknowledgements}
The  research of P.A.  is supported by Sistema Nacional de Investigadores (M\'exico). 

\bibliography{local}

\end{document}